\documentclass[aps,pre,preprint,superscriptaddress,showpacs]{revtex4-1}

\usepackage{amsmath, amsfonts, amsthm, amssymb, mathrsfs, enumerate, graphicx, cases, hyperref}
\usepackage[T1]{fontenc}

\newcommand{\R}{{\mathbb R}}

\begin{document}


\title{Behaviour of asymptotically electro-$\Lambda$ spacetimes}


\author{Vee-Liem Saw}
\email[]{VeeLiem@maths.otago.ac.nz}
\affiliation{Department of Mathematics and Statistics, University of Otago, Dunedin 9016, New Zealand}


\date{\today}

\begin{abstract}
We present the asymptotic solutions for spacetimes with non-zero cosmological constant $\Lambda$ coupled to Maxwell fields, using the Newman-Penrose formalism. This extends a recent work that dealt with the vacuum Einstein (Newman-Penrose) equations with $\Lambda\neq0$. The results are given in two different null tetrads: the Newman-Unti and Szabados-Tod null tetrads, where the peeling property is exhibited in the former but not the latter. Using these asymptotic solutions, we discuss the mass-loss of an isolated electro-gravitating system with cosmological constant. In a universe with $\Lambda>0$, the physics of electromagnetic (EM) radiation is relatively straightforward compared to those of gravitational radiation: 1) It is clear that outgoing EM radiation results in a decrease to the Bondi mass of the isolated system. 2) It is also perspicuous that if any incoming EM radiation from elsewhere is present, those beyond the isolated system's cosmological horizon would eventually arrive at the spacelike $\mathcal{I}$ and increase the Bondi mass of the isolated system. Hence, the (outgoing and incoming) EM radiation fields do not couple with $\Lambda$ in the Bondi mass-loss formula in an unusual manner, unlike the gravitational counterpart where outgoing gravitational radiation induces non-conformal flatness of $\mathcal{I}$. These asymptotic solutions to the Einstein-Maxwell-de Sitter equations presented here may be used to extend a raft of existing results based on Newman-Unti's asymptotic solutions to the Einstein-Maxwell equations where $\Lambda=0$, to now incorporate the cosmological constant $\Lambda$.
\end{abstract}


\maketitle


\section{Introduction}

The behaviour of asymptotically flat empty spacetimes was first described by Newman and Unti \cite{newunti62}, by solving the Newman-Penrose equations \cite{newpen62} (which are equivalent to the Einstein field equations). The Newman-Penrose formalism \cite{newpen62,Pen87,Pen88,don} was born out of the use of null tetrads and complex functions --- leading to great simplifications in many places, as well as being useful in dealing with problems involving radiation. This formalism can be defined in terms of 2-spinors, thus introducing spinorial quantities underlying the usual tensorial objects \cite{pen60}. Certain proofs that have been at best tedious by tensorial means turn out to be relatively straightforward using the spinorial counterpart. For instance, the Petrov classification of the Weyl curvature can be formulated in a perspicuous manner using the complex Weyl spinor. In fact, Newman-Penrose's original paper provided a rather quick proof of the Goldberg-Sachs theorem as an immediate application of their newly-enunciated formalism \cite{newpen62} (reproduced in Ref. \cite{don}), as well as illustrating the peeling property of the (vacuum) curvature tensor that was established by Sachs \cite{Sachs61}. In Newman-Unti's paper itself \cite{newunti62} where the asymptotic solutions to the vacuum Newman-Penrose equations are derived (with $\Lambda=0$), they showed how the Bianchi identity involving the derivative along future null infinity $\mathcal{I}$ of the dyad component of the Weyl spinor $\Psi_2$ leads to the Bondi mass-loss formula \cite{Bondi60,Bondi62}. For an extensive coverage of the related history, see Ref. \cite{Fra04}. Concise descriptions on the Petrov classification, Goldberg-Sachs theorem, null and orthonormal tetrads, spin coefficients, Newman-Penrose formalism, etc. may also be found in Ref. \cite{Ste}, though their conventions differ from those in Refs. \cite{Pen87,Pen88,don}.

A recent work extended this to include the cosmological constant $\Lambda$, motivated by the problem of generalising the Bondi mass and obtaining the mass-loss formula for an isolated gravitating system due to energy carried away by gravitational waves in a de Sitter background universe \cite{Vee2016}. Intense research has been devoted over the past few years on this problem \cite{Vee2016,Szabados,ash1,ash2,ash3,ash4,chi1,chi2,Chrusciel,gracos1,gracos2,Vee2017b} as we now know by observations \cite{cosmo1,cosmo2} that our universe expands at an accelerated rate. A positive cosmological constant $\Lambda>0$ in the Einstein field equations provides the simplest explanation that accounts for such an accelerated expansion rate.

The purpose of this paper is to present the results for the asymptotic solutions to the Einstein-de Sitter equations coupled with the Maxwell fields. The setup and derivation are essentially following those already developed extensively in Ref. \cite{Vee2016} for the $\Lambda$-vacuum case. Here, there are four additional Maxwell equations, with the spin coefficient equations and the Bianchi identities now coupled to the Maxwell fields $\phi_a$, via the Ricci spinor $\Phi_{ab}=k\phi_a\bar{\phi}_b$, where $k=2G/c^4$ (see appendix in Ref. \cite{newpen62}). As shown in Ref. \cite{Vee2016}, the assumption on the fall-off for $\Psi_0=O(r^{-5})$ leads to the peeling property in spacetimes with a cosmological constant \footnote{The cosmological constant is well, a constant. In the Newman-Penrose formalism, the peeling property can be shown by assuming $\Psi_0=O(r^{-5})$ and using the Bianchi identities. This was worked out for the asymptotically flat case in Ref. \cite{newpen62}. What appears in the Bianchi identities involving the Ricci scalar (which is proportional to the cosmological constant) are only its derivatives, not itself --- which are all zero since it is a constant. Thus, one still arrives at the peeling property with a cosmological constant present, as it has no effect on the Bianchi identities \cite{Vee2016}. A different proof involving the use of spinors for the peeling property of (weakly) asymptotically simple spacetimes (which include asymptotically de Sitter spacetimes) may be found in Section 9.7 of Ref. \cite{Pen88}.}. Similarly, the dyad component of the Maxwell spinor $\phi_0$ is taken to fall off as $O(r^{-3})$, giving a corresponding peeling property for the Maxwell spinor. To make this paper self-contained, we recall the necessary background from Ref. \cite{Vee2016} in the next section: the use of spherical coordinates with a retarded null coordinate $u$, the Newman-Unti null tetrad, the fall-offs of the spin coefficients and various quantities defining the null tetrad, as well as the structure of $\mathcal{I}$. Our choice of using these spherical coordinates is perhaps crucial, because they cover the entire de Sitter spacetime \cite{GP} and \emph{naturally reduce to the asymptotically flat case \cite{newunti62,Pen88} when $\Lambda=0$}. This is hence a fitting generalisation to include any $\Lambda\in\R$.

Apart from this Newman-Unti (NU) null tetrad, we also give the results using the Szabados-Tod (ST) null tetrad \cite{Szabados} \footnote{We should clarify that when we say ``Szabados-Tod null tetrad'', we are referring to a null tetrad comprising their $\vec{l}$ and $\vec{n}$, whilst retaining Newman-Unti's $\vec{m}$ and $\vec{\bar{m}}$. This will be defined explicitly in Section 2.}. The latter is defined based on a conformal rescaling which is symmetric with respect to the spin frame $o^A$ and $\iota^A$, as they do not distinguish between the outgoing $\vec{l}$ and incoming $\vec{n}$ null vectors when discussing the conformally rescaled spacetime with spacelike $\mathcal{I}$. In the ST null tetrad, the dyad components of the Weyl and Maxwell spinors do not exhibit the peeling property. Instead, by assuming $\Psi_0=O(r^{-3})$ and $\phi_0=O(r^{-2})$, we then find that $\Psi_i=O(r^{-3})$, $i=1, 2, 3, 4$ and $\phi_j=O(r^{-2})$, $j=1, 2$ \footnote{In Ref. \cite{Pen88}, the proof of the peeling property requires that $\vec{l}$ is related to an affine parameter $r$: $l^a\nabla_ar=1$ (see their Eq. (9.7.3)). The NU null tetrad vector $\vec{l}=\vec{\partial}_r$ satisfies this, but the ST null tetrad vector $\vec{l}=r\vec{\partial}_r$ does not. For the latter, $r$ is not an affine parameter, which is reflected by the non-zero $\gamma'=-1/2$ (see summary in Section 3B).}. This should not be surprising, since a rescaling of the null vectors $\vec{l}$ and $\vec{n}$ (whilst preserving their orthonormalisation properties) actually corresponds to a boost transformation which affects the fall-offs of various quantities (see Appendix A). The ST null tetrad is just one such boost transformation of the NU null tetrad. We would like to point out that in the NU null tetrad, the spin coefficients $\rho'$ and $\gamma$ have leading terms of order $O(r)$ (see summary in Section 3) which appear to blow up as $r\rightarrow\infty$. This is nevertheless, a ramification of spacetime being \emph{asymptotically de Sitter}, where in these spherical coordinates we have the inverse metric component $g^{rr}=\Lambda r^2/3+O(1)$, in contrast to just $O(1)$ when $\Lambda=0$. (Even empty de Sitter spacetime itself has $\rho'=-\Lambda r/6+r^{-1}/2$ and $\gamma=-\Lambda r/6$ \cite{Vee2016}.) Alternatively, we see that in the ST null tetrad, $\rho'$ and $\gamma$ are both of order $O(1)$, giving a finite limit at $r\rightarrow\infty$. The summary of the asymptotic solutions for the Einstein-Maxwell-de Sitter theory, in both the NU and ST null tetrads are found in Section 3 \footnote{We derived these asymptotic solutions for both NU and ST null tetrads independently, i.e. working through all 42 Newman-Penrose-Maxwell equations separately on different occasions. Then, we verified that one could be obtained from the other by the appropriate boost transformation, as expected.}.

This summary should prove to be highly useful, as many results for the asymptotically flat case based on these solutions may be extended to include $\Lambda$. For instance, one may study the memory effect following Ref. \cite{framemory}. A derivation of the Bondi mass-loss formula by Frauendiener using an integral formula that arose from the Nester-Witten identity essentially applies these asymptotic solutions with $\Lambda=0$ \cite{fra97}. A most recent paper by Newman brought up an enigma in asymptotically flat spacetimes \cite{Newmanenigma}, where he makes use of the Einstein-Maxwell asymptotic solutions and found perplexing relationships mimicking classical Newtonian and Maxwell theories. Such motions and radiation however, do not occur in physical spacetime --- instead, they live on an associated complex four-dimensional manifold. It is certainly intriguing to subsequently investigate what the cosmological constant has to say, as our universe is described by a small but non-zero $\Lambda>0$. As an application of these newly derived asymptotic solutions to the Einstein-Maxwell-de Sitter theory, we discuss the situation concerning an isolated electro-gravitating system in Section 4 based on the Bianchi identity involving the $u$-derivative of $\Psi_2$.

The notations and conventions used here follow those in Ref. \cite{Vee2016}, which are in accordance to Refs. \cite{Pen87,Pen88}. One may refer to these references for the definitions of the spin coefficients, Weyl spinor, Ricci spinor, etc., or see Appendix B \footnote{Well, for past research employing the asymptotic solutions to the Einstein-Maxwell theory with $\Lambda=0$ based on Ref. \cite{Pen88}, our results here with $\Lambda$ may be directly applied since we follow those conventions. Furthermore, setting $\Lambda=0$ reduces exactly to those in Ref. \cite{Pen88}.}.

\section{Background}

We recapitulate some necessary details that have been elaborated in Ref. \cite{Vee2016}. Our choice of coordinates is motivated by one which holds for any $\Lambda\in\R$. In spherical coordinates, the metric for de Sitter spacetime is \cite{GP}:
\begin{eqnarray}\label{SchwarzdS}
g=J(r)dt^2-\frac{1}{J(r)}dr^2-r^2d\Omega^2,
\end{eqnarray}
where $J(r)=-\Lambda r^2/3+1$, with $\Lambda$ being the cosmological constant ($\Lambda>0$, $\Lambda<0$ and $\Lambda=0$ correspond to the de Sitter, anti-de Sitter and Minkowski spacetimes, respectively). The unit 2-sphere is $d\Omega^2=d\theta^2+\sin^2{\theta}d\phi^2$. These coordinates have the following ranges: $-\infty<t<\infty, 0\leq r<\infty, 0\leq\theta\leq\pi, 0\leq\phi<2\pi$. The metric in Eq. (\ref{SchwarzdS}) can be expressed in terms of a ``retarded null coordinate $u$'' \cite{Vee2016}:
\begin{eqnarray}
g=J(r)du^2+2dudr-r^2d\Omega^2.
\end{eqnarray}
This is the form of the metric upon which the generalisation to asymptotically de Sitter spacetimes is based upon. In particular, a general (Newman-Unti) null tetrad $\vec{l},\vec{n},\vec{m},\vec{\bar{m}}$ where $g^{ab}=l^an^b+n^al^b-m^a\bar{m}^b-\bar{m}^am^b$ is defined as \cite{Vee2016}:
\begin{eqnarray}
\vec{l}&=&\vec{\partial}_r\\
\vec{n}&=&\vec{\partial}_u+U\vec{\partial}_r+X^\mu\vec{\partial}_\mu\\
\vec{m}&=&\omega\vec{\partial}_r+\xi^\mu\vec{\partial}_\mu\\
\vec{\bar{m}}&=&\bar{\omega}\vec{\partial}_r+\overline{\xi^\mu}\vec{\partial}_\mu,
\end{eqnarray}
where $\mu$ denotes the two coordinates $\theta$ and $\phi$. These are labels for the null generators of those outgoing null hypersurfaces $u=$ constant, and in general do not correspond to the usual two spherical angular coordinates. The functions $U(u,r,\theta,\phi)$, $X^\mu(u,r,\theta,\phi)$, $\omega(u,r,\theta,\phi)$, $\xi^\mu(u,r,\theta,\phi)$ are prescribed to have the following expansions in inverse powers of $r$ \footnote{In Ref. \cite{Vee2016} which dealt exclusively with the physical spacetime, the form for $U$ was stipulated by investigating/calculating it for the Schwarzschild-(anti-)de Sitter spacetime: where $U=\Lambda r^2/6-1/2+Mr^{-1}.$ Then, the generalisation to asymptotically (anti-)de Sitter spacetimes was to suppose that $U=\Lambda r^2/6+U^o_0+U^o_1r^{-1}+O(r^{-2})$, having an expansion as inverse powers of $r$ with sufficiently many orders, and with no order $r$ term. (Since the mass $M$ only appears at order $r^{-1}$, it was assumed that it does not affect $U$ at order $r$.) This ansatz was successful in obtaining the general asymptotic solutions for asymptotically (anti-)de Sitter spacetimes, which reduce to those for asymptotically flat spacetimes \cite{Pen88} when $\Lambda$ is set to zero. Of course, it was already known at the time that this is indeed the correct form based on Ref. \cite{Szabados}'s study of the conformal structure.}
\begin{eqnarray}
U&=&\frac{\Lambda}{6}r^2+O(1)\\
X^\mu&=&O(r^{-1})\\
\omega&=&O(r^{-1})\\
\xi^\mu&=&(\xi^\mu)^or^{-1}+O(r^{-2}).\label{defxi}
\end{eqnarray}
The coefficients of each term in the expansions are unknown functions of $u,\theta,\phi$, which were determined by solving the 38 vacuum Newman-Penrose equations \cite{Vee2016}. Here, they will be obtained from those 38 equations with Maxwell fields, plus the 4 Maxwell equations.

Before solving those 42 equations, the fall-offs for the 12 complex spin coefficients are stipulated. Some geometrical properties like $\vec{l}$ being tangent to the outgoing null geodesics, the remaining null tetrad vectors being parallel transported along $\vec{l}$, the freedoms involving the origin of the affine parameter $r$ of those outgoing null geodesics, as well as spatial rotations relating to $\vec{m}$ and $\vec{\bar{m}}$ are applied, and the fall-offs are
\begin{eqnarray}
\kappa&=&0, \gamma'=0, \tau'=0,\\
\kappa'&=&O(1)\\
\sigma&=&O(r^{-1})\\
\sigma'&=&O(1)\\
\tau&=&O(r^{-1})\\
\gamma&=&\gamma^o_{-1}r+O(r^{-2})\\
\rho&=&\rho^o_1r^{-1}+O(r^{-3})\\
\rho'&=&\rho'^o_{-1}r+O(r^{-1})\\
\alpha&=&O(r^{-1})\\
\alpha'&=&O(r^{-1}).
\end{eqnarray}
These were stated by studying the Schwarzschild-de Sitter spacetime (which necessarily implies that $\rho'^o_{-1}$ and $\gamma^o_{-1}$ are non-zero due to $\Lambda\neq0$), as well as to ensure that $\sigma^o$ (which is the $r^{-2}$ term for $\sigma$) is not forced to vanish (and this gives rise to $O(1)$ terms for $\sigma'$ and $\kappa'$). The absence of the terms of orders $O(1)$ and $O(r^{-1})$ in $\gamma$ as well as the absence of an $O(1)$ term in $\rho'$ were assumed since for the Schwarszschild-de Sitter spacetime, $\gamma=-\Lambda r/6+Mr^{-2}/2$ and $\rho'=-\Lambda r/6+r^{-1}/2-Mr^{-2}$, i.e. the mass $M$ does not affect those orders \footnote{All these fall-offs are consistent with the results presented in Ref. \cite{Szabados} who studied the conformal structure of de Sitter-like spacetimes. In other words, these are not just ad hoc assumptions but they correspond to fixing the gauge freedoms (please see Ref. \cite{Vee2016} for more detailed elaborations on this).}. The last piece of required information is to have the fall-off for the dyad component of the Weyl spinor $\Psi_0$ being $O(r^{-5})$, in accordance to the asymptotically flat case \cite{newpen62,newunti62}.

Unlike the case for asymptotically flat spacetimes, the presence of a non-zero cosmological constant generally forbids the leading order terms $(\xi^{\mu})^o$ in Eqs. (\ref{defxi}) to be specified in a manner corresponding to a round unit sphere \cite{Szabados}. This is because that would lead to a conformally flat $\mathcal{I}$ with the repercussion that gravitational waves do not carry energy away from the isolated source \cite{ash1}. In the Newman-Penrose formalism, this is reflected by the pair of metric equations involving $D'\xi^\mu$, which yield \cite{Vee2016} (see also the summary in section 3)
\begin{eqnarray}\label{xi}
(\dot{\xi}^\mu)^o=-\frac{\Lambda}{3}\sigma^o(\overline{\xi^\mu})^o.
\end{eqnarray}
Well, $\mathcal{I}$ can be thought of as being foliated by these 2-surfaces of constant $u$ (see the setup of $\mathcal{I}$ with $\Lambda>0$ in Ref. \cite{Szabados} that describes this). The above equations mean that whilst one may pick one such 2-surface on $\mathcal{I}$ to be a round sphere for some $u=u_0$, a different $u$ would generally have a \emph{topological 2-sphere} when $\Lambda\neq0$ \emph{and} $\sigma^o\neq0$. Furthermore, the Gauss curvature for this topological 2-sphere is
\begin{eqnarray}
K=1+\frac{2\Lambda}{3}\int{\textrm{Re}(\eth^2\bar{\sigma}^o)du},
\end{eqnarray}
obtained from the spin coefficient equation involving the derivatives of $\alpha$ and $\alpha'$ intrinsic to those 2-surfaces \cite{Vee2016,Pen88}. The $\eth$ operator is defined in an expected manner and acts on spin-weighted quantities (see Ref. \cite{Vee2016} and/or the summary in the next section). When $\sigma^o=0$, then $\mathcal{I}$ is conformally flat --- its Cotton-York tensor is zero. With the right hand side of Eqs. (\ref{xi}) being 0 (due to $\sigma^o=0$), the choice of $(\xi^o)^\mu$ being a round sphere at some $u=u_0$ would remain even as $u$ varies. Also, we see that the Gauss curvature is then just 1.

For an axisymmetric system (i.e. no explicit $\phi$ dependence), Eqs. (\ref{xi}) may be solved explicitly. With $3f(u,\theta):=\int{\sigma^o(u,\theta)du}$ where $\sigma^o(u,\theta)$ is real \footnote{Well, the two polarisation modes of gravitational waves are encoded into the real and imaginary parts of $\sigma^o$. With axisymmetry, there is only one polarisation mode \cite{ash4} and we can thus specify $\sigma^o(u,\theta)$ to be a real function.}, we find that
\begin{eqnarray}
(\xi^\theta)^o&=&\frac{1}{\sqrt{2}}e^{-\Lambda f(u,\theta)}\\
(\xi^\phi)^o&=&\frac{i}{\sqrt{2}}e^{\Lambda f(u,\theta)}\csc{\theta},
\end{eqnarray}
satisfy that pair of equations, and so the metric for the 2-surface of constant $u$ on $\mathcal{I}$ is
\begin{eqnarray}\label{2axi}
g_{2,axi}=e^{2\Lambda f(u,\theta)}d\theta^2+e^{-2\Lambda f(u,\theta)}\sin^2{\theta}d\phi^2.
\end{eqnarray}
Notice that this is not a round unit sphere, unless $\Lambda=0$ (i.e. spacetime is asymptotically flat), or $f=0$ (so $\sigma^o=0$).

The full general (without axisymmetry) $(3+1)$-$d$ spacetime would have the metric
\begin{eqnarray}
g=-\frac{\Lambda}{3}r^2du^2-r^2g_2+O(r),
\end{eqnarray}
where $g_2$ is the metric for the topological 2-sphere constructed from $(\xi^\mu)^o$, and $O(r)$ represents the collection of terms of higher order than $r^2$, i.e. $r^{-2}$ times those terms represented by $O(r)$ would go to zero as $r\rightarrow\infty$. Incidentally, this agrees with the form of the physical metric for de Sitter-like spacetimes given by Eq. (4.18) in Ref. \cite{Szabados}, obtained by studying the conformal structure. Now, we can define a conformally rescaled metric $\tilde{g}:=C^2g$, where $C=r^{-1}$ is the conformal factor. Then,
\begin{eqnarray}
\tilde{g}=-\frac{\Lambda}{3}du^2-g_2+O(r^{-1}).
\end{eqnarray}
In the limit where $r\rightarrow\infty$, this gives the metric for null infinity $\mathcal{I}$ \footnote{Recall that the hypersurfaces $u=$ constant are outgoing null cones, with $r$ being an affine parameter of the null geodesics generating these null cones. So $r\rightarrow\infty$ would reach infinity along a null light ray, which is null infinity --- since by definition, null infinity is the hypersurface where light rays would eventually end up after infinite physical time.}
\begin{eqnarray}
\tilde{g}=-\frac{\Lambda}{3}du^2-g_2.
\end{eqnarray}
With axisymmetry, we have $g_2=g_{2,axi}$ in Eq. (\ref{2axi}) so
\begin{eqnarray}\label{scriaxi}
\tilde{g}=-\frac{\Lambda}{3}du^2-e^{2\Lambda f(u,\theta)}d\theta^2-e^{-2\Lambda f(u,\theta)}\sin^2{\theta}d\phi^2.
\end{eqnarray}
We see that the nature of the hypersurface $\mathcal{I}$ depends on the cosmological constant: It is a spacelike hypersurface for $\Lambda>0$, a timelike hypersurface for $\Lambda<0$, and degenerates into a null hypersurface for $\Lambda=0$. The axisymmetric metric for $\mathcal{I}$ given by Eq. (\ref{scriaxi}) incidentally, is also found in Ref. \cite{chi1} who studied this problem of gravitational radiation with a cosmological constant by employing Bondi et al.'s original approach \cite{Bondi62} in enunciating an axisymmetric metric ansatz. To account for the cosmological constant, Ref. \cite{chi1} introduced a new leading term in the function $\gamma$ (which when prescribed for some $u$, allows for other unknowns to be solved systematically via the Einstein field equations). This new leading term in their ansatz would also lead to a non-conformally flat $\mathcal{I}$, i.e. Eq. (\ref{scriaxi}).

As mentioned and elaborated in the previous section, we also give the asymptotic solutions using a ``Szabados-Tod''-type null tetrad \cite{Szabados}, where $\vec{l}\rightarrow r\vec{l}$, $\vec{n}\rightarrow\vec{n}/r$. The same (NU) $\vec{m},\vec{\bar{m}}$ are used to complete the ST tetrad of null vectors \footnote{The $\vec{m},\vec{\bar{m}}$ in Ref. \cite{Szabados} do not have a $\vec{\partial}_r$ component.}.

The approach in Refs. \cite{newpen62,newunti62} begins by considering a family of null hypersurfaces $u=\ $constant opening up towards the future. A null vector field $\vec{l}$ is defined as $\vec{l}=\vec{\nabla}u$, with $\vec{l}$ also satisfying $l^a\nabla_a l^b=0$. This $\vec{l}$ induces a congruence of null geodesics tangent to these null hypersurfaces, and so there is an associated affine parameter $r$ for these null geodesics. With this $\vec{l}$ as well as the $u$ and $r$ coordinates, two angular labels $\theta,\phi$ can be defined, and the null tetrad subsequently completed with $\vec{n}, \vec{m}, \vec{\bar{m}}$ (obeying the orthonormality conditions). Finally the metric is obtained from the null tetrad, expressed in the ${u, r, \theta, \phi}$ coordinates.

On the other hand, the way we \cite{Vee2016} go about with the setup is essentially coming from the other direction. We are looking for a form/set of coordinates for the metric which are ``useful'': such that the metric turns out to be of the form constructed in Refs. \cite{newpen62,newunti62}, and reduces to the asymptotically flat case when $\Lambda=0$. This can be achieved with spherical coordinates as presented here. From this point of view, the coordinate $r$ may be regarded as \emph{the usual spherical radial coordinate}. With this as the definition of $r$, one is free to define $\vec{l}$, which induces a congruence of null geodesics such that this $r$ happens to: (1) be an affine parameter (NU); (2) not be an affine parameter (ST). \emph{This argument carries over to the asymptotically flat case, since it holds when $\Lambda=0$}.

\section{Summary}

\subsection{Newman-Unti null tetrad}

Asymptotically electro-$\Lambda$ spacetimes may be described by the following NU null tetrad \cite{newunti62}:
\begin{eqnarray}
\vec{l}&=&\vec{\partial}_r\\
\vec{n}&=&\vec{\partial}_u+U\vec{\partial}_r+X^\theta\vec{\partial}_\theta+X^\phi\vec{\partial}_\phi\\
\vec{m}&=&\omega\vec{\partial}_r+\xi^\theta\vec{\partial}_\theta+\xi^\phi\vec{\partial}_\phi\\
\vec{\bar{m}}&=&\bar{\omega}\vec{\partial}_r+\overline{\xi^\theta}\vec{\partial}_\theta+\overline{\xi^\phi}\vec{\partial}_\phi,
\end{eqnarray}
where
\begin{eqnarray}
U&=&\frac{\Lambda}{6}r^2-\left(\frac{1}{2}K+\frac{\Lambda}{2}|\sigma^o|^2\right)-\textrm{Re}(\Psi^o_2)r^{-1}+O(r^{-2})\\
X^\mu&=&\frac{1}{3}\textrm{Re}(\bar{\Psi}^o_1(\xi^\mu)^o)r^{-3}+O(r^{-4})\\
\omega&=&\eth'\sigma^or^{-1}-\left(\sigma^o\eth\bar{\sigma}^o+\frac{1}{2}\Psi^o_1\right)r^{-2}+O(r^{-3})\\
\xi^\mu&=&(\xi^\mu)^or^{-1}-\sigma^o(\overline{\xi^\mu})^or^{-2}+|\sigma^o|^2(\xi^\mu)^or^{-3}+\left(\frac{1}{6}\Psi^o_0-\sigma^o|\sigma^o|^2\right)(\overline{\xi^\mu})^or^{-4}+O(r^{-5}).
\end{eqnarray}

The components of the inverse metric are related to the null tetrad via $g^{ab}=l^an^b+n^al^b-m^a\bar{m}^b-\bar{m}^am^b$:
\begin{eqnarray}
g^{ab}&=&
\begin{pmatrix}
  0 & 1 & 0 & 0 \\
  1 & 2(U-|\omega|^2) & X^\theta-2\textrm{Re}(\xi^\theta\bar{\omega}) & X^\phi-2\textrm{Re}(\xi^\phi\bar{\omega}) \\
  0 & X^\theta-2\textrm{Re}(\xi^\theta\bar{\omega}) & -2|\xi^\theta|^2 & -2\text{Re}(\xi^\theta\overline{\xi^\phi}) \\
	0 & X^\phi-2\textrm{Re}(\xi^\phi\bar{\omega}) & -2\text{Re}(\xi^\theta\overline{\xi^\phi}) & -2|\xi^\phi|^2
\end{pmatrix}\\
&=&
\begin{pmatrix}
  0 & 1 & 0 & 0 \\
  1 & \frac{\Lambda}{3}r^2 & -2\textrm{Re}((\xi^\theta)^o\eth\bar{\sigma}^o)r^{-2} & -2\textrm{Re}((\xi^\phi)^o\eth\bar{\sigma}^o)r^{-2} \\
  0 & -2\textrm{Re}((\xi^\theta)^o\eth\bar{\sigma}^o)r^{-2} & -2|(\xi^\theta)^o|^2r^{-2} & -2\text{Re}((\xi^\theta)^o(\overline{\xi^\phi})^o)r^{-2} \\
	0 & -2\textrm{Re}((\xi^\phi)^o\eth\bar{\sigma}^o)r^{-2} & -2\text{Re}((\xi^\theta)^o(\overline{\xi^\phi})^o)r^{-2} & -2|(\xi^\phi)^o|^2r^{-2}
\end{pmatrix}\nonumber\\
&&+
\begin{pmatrix}
  0 & 0 & 0 & 0 \\
  0 & O(1) & O(r^{-3}) & O(r^{-3}) \\
  0 & O(r^{-3}) & O(r^{-3}) & O(r^{-3}) \\
	0 & O(r^{-3}) & O(r^{-3}) & O(r^{-3})\label{metSU}
\end{pmatrix}.
\end{eqnarray}

The spin coefficients are:
\begin{eqnarray}
\kappa&=&0, \gamma'=0, \tau'=0\\
\rho&=&-r^{-1}-|\sigma^o|^2r^{-3}+\left(\frac{1}{3}\textrm{Re}(\bar{\sigma}^o\Psi^o_0)-|\sigma^o|^4-\frac{1}{3}k|\phi^o_0|\right)r^{-5}+O(r^{-6})\\
\sigma&=&\sigma^or^{-2}+\left(\sigma^o|\sigma^o|^2-\frac{1}{2}\Psi^o_0\right)r^{-4}-\frac{1}{3}\Psi^1_0r^{-5}+O(r^{-6})\\
\rho'&=&-\frac{\Lambda}{6}r+\left(\frac{1}{2}K+\frac{\Lambda}{3}|\sigma^o|^2\right)r^{-1}+(\Psi^o_2+\sigma^o\dot{\bar{\sigma}}^o)r^{-2}+O(r^{-3})\\
\sigma'&=&-\frac{\Lambda}{6}\bar{\sigma}^o-\dot{\bar{\sigma}}^or^{-1}-\left(\frac{1}{2}K\bar{\sigma}^o+\frac{\Lambda}{3}\bar{\sigma}^o|\sigma^o|^2+\frac{\Lambda}{12}\bar{\Psi}^o_0\right)r^{-2}+O(r^{-3})\\
\tau&=&-\frac{1}{2}\Psi^o_1r^{-3}+O(r^{-4})\\
\gamma&=&-\frac{\Lambda}{6}r-\frac{1}{2}\Psi^o_2r^{-2}+O(r^{-3})\\
\kappa'&=&\frac{\Lambda}{3}\eth\bar{\sigma}^o+\left(\Psi^o_3-\frac{\Lambda}{12}\bar{\Psi}^o_1\right)r^{-1}+O(r^{-2})\\
\alpha&=&\alpha^or^{-1}+\bar{\alpha}^o\bar{\sigma}^or^{-2}+\alpha^o|\sigma^o|^2r^{-3}+O(r^{-4})\\
\alpha'&=&\bar{\alpha}^or^{-1}+\alpha^o\sigma^or^{-2}+\left(\bar{\alpha}^o|\sigma^o|^2+\frac{1}{2}\Psi^o_1\right)r^{-3}+O(r^{-4}).
\end{eqnarray}

The Weyl spinor has dyad components:
\begin{eqnarray}
\Psi_0&=&\Psi^o_0r^{-5}+\Psi^1_0r^{-6}+O(r^{-7})\\
\Psi_1&=&\Psi^o_1r^{-4}-\left(\eth'\Psi^o_0-3k\phi^o_0\bar{\phi}^o_1\right)r^{-5}+O(r^{-6})\\
\Psi_2&=&\Psi^o_2r^{-3}-\left(\eth'\Psi^o_1-2k|\phi^o_1|^2-\frac{\Lambda}{6}\bar{\sigma}^o\Psi^o_0+\frac{k\Lambda}{6}|\phi^o_0|^2\right)r^{-4}+O(r^{-5})\\
\Psi_3&=&\Psi^o_3r^{-2}-\left(\eth'\Psi^o_2-k\phi^o_2\bar{\phi}^o_1-\frac{\Lambda}{3}\bar{\sigma}^o\Psi^o_1+\frac{k\Lambda}{3}\phi^o_1\bar{\phi}^o_0\right)r^{-3}+O(r^{-4})\\
\Psi_4&=&\Psi^o_4r^{-1}-\left(\eth'\Psi^o_3-\frac{\Lambda}{2}\bar{\sigma}^o\Psi^o_2+\frac{k\Lambda}{2}\phi^o_2\bar{\phi}^o_0\right)r^{-2}+O(r^{-3}),
\end{eqnarray}
with the Bianchi identities:
\begin{eqnarray}
\dot{\Psi}^o_0&=&\eth\Psi^o_1+3\sigma^o\Psi^o_2+3k\phi^o_0\bar{\phi}^o_2+\frac{\Lambda}{6}\Psi^1_0\label{Biazero}\\
\dot{\Psi}^o_1&=&\eth\Psi^o_2+2\sigma^o\Psi^o_3+2k\phi^o_1\bar{\phi}^o_2-\frac{\Lambda}{6}\eth'\Psi^o_0+\frac{k\Lambda}{3}\phi^o_0\bar{\phi}^o_1\\
-\frac{\partial}{\partial u}(\Psi^o_2+\sigma^o\dot{\bar{\sigma}}^o)&=&-|\dot{\sigma}^o|^2-\eth\Psi^o_3-k|\phi^o_2|^2-\frac{\Lambda}{3}K|\sigma^o|^2+\frac{\Lambda}{3}\sigma^o\eth'\eth\bar{\sigma}^o+\frac{\Lambda}{6}\eth'\Psi^o_1-\frac{2\Lambda^2}{9}|\sigma^o|^4\nonumber\\&&-\frac{\Lambda^2}{18}\textrm{Re}(\bar{\sigma}^o\Psi^o_0)+\frac{k\Lambda^2}{36}|\phi^o_0|^2.\label{Biatwo}
\end{eqnarray}

Also:
\begin{eqnarray}
&&\textrm{Im}(\Psi^o_2+\sigma^o\dot{\bar{\sigma}}^o+\eth^2\bar{\sigma}^o)=0\\
&&\Psi^o_3=-\eth\dot{\bar{\sigma}}^o-\frac{\Lambda}{6}\bar{\Psi}^o_1+\frac{\Lambda}{3}\sigma^o\eth'\bar{\sigma}^o+\eth'U^o_0\\
&&\Psi^o_4=-\ddot{\bar{\sigma}}^o+\frac{\Lambda}{3}K\bar{\sigma}^o-\frac{\Lambda}{3}\eth'\eth\bar{\sigma}^o+\frac{2\Lambda^2}{9}\bar{\sigma}^o|\sigma^o|^2+\frac{\Lambda^2}{36}\bar{\Psi}^o_0\label{psi4}\\
&&\dot{\alpha}^o=\frac{\Lambda}{3}\bar{\alpha}^o\bar{\sigma}^o+\frac{\Lambda}{6}\eth\bar{\sigma}^o\\
&&(\dot{\xi}^\mu)^o=-\frac{\Lambda}{3}\sigma^o(\overline{\xi^\mu})^o\\
&&\textrm{Im}(\eth'(\xi^\mu)^o)=0\\
&&K=1+\frac{2\Lambda}{3}\int{\text{Re}(\eth^2\bar{\sigma}^o)du}\label{K},
\end{eqnarray}
where $U^o_0=-\frac{1}{2}K-\frac{\Lambda}{2}|\sigma^o|^2$, and $K$ is the Gauss curvature of the compact 2-surface of constant $u$ on $\mathcal{I}$.

The dyad components of the Maxwell spinor are:
\begin{eqnarray}
\phi_0&=&\phi^o_0r^{-3}+\phi^1_0r^{-4}+O(r^{-5})\\
\phi_1&=&\phi^o_1r^{-2}-\eth'\phi^o_0r^{-3}+O(r^{-4})\\
\phi_2&=&\phi^o_2r^{-1}-\left(\eth'\phi^o_1-\frac{\Lambda}{6}\bar{\sigma}^o\phi^o_0\right)r^{-2}+O(r^{-3}),
\end{eqnarray}
with Maxwell equations:
\begin{eqnarray}
\dot{\phi}^o_0&=&\eth\phi^o_1+\sigma^o\phi^o_2+\frac{\Lambda}{6}\phi^1_0\label{max1}\\
\dot{\phi}^o_1&=&\eth\phi^o_2-\frac{\Lambda}{6}\eth'\phi^o_0.\label{max2}
\end{eqnarray}


For both NU and ST null tetrads, the $\eth$ and $\eth'$ operators acting on a scalar $\eta$ with spin-weight $s$ are defined as \cite{Pen87}:
\begin{eqnarray}
\eth\eta:=(\delta^o+2s\bar{\alpha}^o)\eta, \eth':=(\delta'^o-2s\alpha^o)\eta,
\end{eqnarray}
where $\displaystyle\delta^o:=(\xi^\theta)^o\frac{\partial}{\partial\theta}+(\xi^\phi)^o\frac{\partial}{\partial\phi}$ and $\displaystyle\delta'^o:=(\overline{\xi^\theta})^o\frac{\partial}{\partial\theta}+(\overline{\xi^\phi})^o\frac{\partial}{\partial\phi}$.

\subsection{Szabados-Tod null tetrad}

Asymptotically electro-$\Lambda$ spacetimes may be described by the following ST null tetrad \cite{Szabados}:
\begin{eqnarray}
\vec{l}&=&r\vec{\partial}_r\\
\vec{n}&=&\frac{1}{r}\vec{\partial}_u+U\vec{\partial}_r+X^\theta\vec{\partial}_\theta+X^\phi\vec{\partial}_\phi\\
\vec{m}&=&\omega\vec{\partial}_r+\xi^\theta\vec{\partial}_\theta+\xi^\phi\vec{\partial}_\phi\\
\vec{\bar{m}}&=&\bar{\omega}\vec{\partial}_r+\overline{\xi^\theta}\vec{\partial}_\theta+\overline{\xi^\phi}\vec{\partial}_\phi,
\end{eqnarray}
where
\begin{eqnarray}
U&=&\frac{\Lambda}{6}r-\left(\frac{1}{2}K+\frac{\Lambda}{2}|\sigma^o|^2\right)r^{-1}-\textrm{Re}(\Psi^o_2)r^{-2}+O(r^{-3})\\
X^\mu&=&\frac{1}{3}\textrm{Re}(\bar{\Psi}^o_1(\xi^\mu)^o)r^{-4}+O(r^{-5})\\
\omega&=&\eth'\sigma^or^{-1}-\left(\sigma^o\eth\bar{\sigma}^o+\frac{1}{2}\Psi^o_1\right)r^{-2}+O(r^{-3})\\
\xi^\mu&=&(\xi^\mu)^or^{-1}-\sigma^o(\overline{\xi^\mu})^or^{-2}+|\sigma^o|^2(\xi^\mu)^or^{-3}+\left(\frac{1}{6}\Psi^o_0-\sigma^o|\sigma^o|^2\right)(\overline{\xi^\mu})^or^{-4}+O(r^{-5}).
\end{eqnarray}

The components of the inverse metric are related to the null tetrad via $g^{ab}=l^an^b+n^al^b-m^a\bar{m}^b-\bar{m}^am^b$:
\begin{eqnarray}
g^{ab}&=&
\begin{pmatrix}
  0 & 1 & 0 & 0 \\
  1 & 2(rU-|\omega|^2) & rX^\theta-2\textrm{Re}(\xi^\theta\bar{\omega}) & rX^\phi-2\textrm{Re}(\xi^\phi\bar{\omega}) \\
  0 & rX^\theta-2\textrm{Re}(\xi^\theta\bar{\omega}) & -2|\xi^\theta|^2 & -2\text{Re}(\xi^\theta\overline{\xi^\phi}) \\
	0 & rX^\phi-2\textrm{Re}(\xi^\phi\bar{\omega}) & -2\text{Re}(\xi^\theta\overline{\xi^\phi}) & -2|\xi^\phi|^2
\end{pmatrix}\\
&=&\textrm{Eq. }(\ref{metSU}).
\end{eqnarray}

The spin coefficients are:
\begin{eqnarray}
\kappa&=&0, \tau'=0, \gamma'=-\frac{1}{2}\\
\rho&=&-1-|\sigma^o|^2r^{-2}+\left(\frac{1}{3}\textrm{Re}(\bar{\sigma}^o\Psi^o_0)-|\sigma^o|^4-\frac{1}{3}k|\phi^o_0|\right)r^{-4}+O(r^{-5})\\
\sigma&=&\sigma^or^{-1}+\left(\sigma^o|\sigma^o|^2-\frac{1}{2}\Psi^o_0\right)r^{-3}-\frac{1}{3}\Psi^1_0r^{-4}+O(r^{-5})\\
\rho'&=&-\frac{\Lambda}{6}+\left(\frac{1}{2}K+\frac{\Lambda}{3}|\sigma^o|^2\right)r^{-2}+(\Psi^o_2+\sigma^o\dot{\bar{\sigma}}^o)r^{-3}+O(r^{-4})\\
\sigma'&=&-\frac{\Lambda}{6}\bar{\sigma}^or^{-1}-\dot{\bar{\sigma}}^or^{-2}-\left(\frac{1}{2}K\bar{\sigma}^o+\frac{\Lambda}{3}\bar{\sigma}^o|\sigma^o|^2+\frac{\Lambda}{12}\bar{\Psi}^o_0\right)r^{-3}+O(r^{-4})\\
\tau&=&-\frac{1}{2}\Psi^o_1r^{-3}+O(r^{-4})\\
\gamma&=&-\frac{\Lambda}{12}-\left(\frac{1}{4}K+\frac{\Lambda}{4}|\sigma^o|^2\right)r^{-2}-\left(\frac{3}{4}\Psi^o_2+\frac{1}{4}\bar{\Psi}^o_2\right)r^{-3}+O(r^{-4})\\
\kappa'&=&\frac{\Lambda}{3}\eth\bar{\sigma}^or^{-2}+\left(\Psi^o_3-\frac{\Lambda}{12}\bar{\Psi}^o_1\right)r^{-3}+O(r^{-4})\\
\alpha&=&\alpha^or^{-1}+\left(\bar{\alpha}^o\bar{\sigma}^o+\frac{1}{2}\eth\bar{\sigma}^o\right)r^{-2}+\left(\alpha^o|\sigma^o|^2-\frac{1}{2}\bar{\sigma}^o\eth'\sigma^o-\frac{1}{4}\bar{\Psi}^o_1\right)r^{-3}+O(r^{-4})\\
\alpha'&=&\bar{\alpha}^or^{-1}+\left(\alpha^o\sigma^o-\frac{1}{2}\eth'\sigma^o\right)r^{-2}+\left(\bar{\alpha}^o|\sigma^o|^2+\frac{1}{2}\sigma^o\eth\bar{\sigma}^o+\frac{3}{4}\Psi^o_1\right)r^{-3}+O(r^{-4}).
\end{eqnarray}

The Weyl spinor has dyad components:
\begin{eqnarray}
\Psi_0&=&\Psi^o_0r^{-3}+\Psi^1_0r^{-4}+O(r^{-5})\\
\Psi_1&=&\Psi^o_1r^{-3}-\left(\eth'\Psi^o_0-3k\phi^o_0\bar{\phi}^o_1\right)r^{-4}+O(r^{-5})\\
\Psi_2&=&\Psi^o_2r^{-3}-\left(\eth'\Psi^o_1-2k|\phi^o_1|^2-\frac{\Lambda}{6}\bar{\sigma}^o\Psi^o_0+\frac{k\Lambda}{6}|\phi^o_0|^2\right)r^{-4}+O(r^{-5})\\
\Psi_3&=&\Psi^o_3r^{-3}-\left(\eth'\Psi^o_2-k\phi^o_2\bar{\phi}^o_1-\frac{\Lambda}{3}\bar{\sigma}^o\Psi^o_1+\frac{k\Lambda}{3}\phi^o_1\bar{\phi}^o_0\right)r^{-4}+O(r^{-5})\\
\Psi_4&=&\Psi^o_4r^{-3}-\left(\eth'\Psi^o_3-\frac{\Lambda}{2}\bar{\sigma}^o\Psi^o_2+\frac{k\Lambda}{2}\phi^o_2\bar{\phi}^o_0\right)r^{-4}+O(r^{-5}),
\end{eqnarray}
with the Bianchi identities:
\begin{eqnarray}
\dot{\Psi}^o_0&=&\eth\Psi^o_1+3\sigma^o\Psi^o_2+3k\phi^o_0\bar{\phi}^o_2+\frac{\Lambda}{6}\Psi^1_0\label{BiazeroST}\\
\dot{\Psi}^o_1&=&\eth\Psi^o_2+2\sigma^o\Psi^o_3+2k\phi^o_1\bar{\phi}^o_2-\frac{\Lambda}{6}\eth'\Psi^o_0+\frac{k\Lambda}{3}\phi^o_0\bar{\phi}^o_1\\
-\frac{\partial}{\partial u}(\Psi^o_2+\sigma^o\dot{\bar{\sigma}}^o)&=&-|\dot{\sigma}^o|^2-\eth\Psi^o_3-k|\phi^o_2|^2-\frac{\Lambda}{3}K|\sigma^o|^2+\frac{\Lambda}{3}\sigma^o\eth'\eth\bar{\sigma}^o+\frac{\Lambda}{6}\eth'\Psi^o_1-\frac{2\Lambda^2}{9}|\sigma^o|^4\nonumber\\&&-\frac{\Lambda^2}{18}\textrm{Re}(\bar{\sigma}^o\Psi^o_0)+\frac{k\Lambda^2}{36}|\phi^o_0|^2.\label{BiatwoST}
\end{eqnarray}

Also:
\begin{eqnarray}
&&\textrm{Im}(\Psi^o_2+\sigma^o\dot{\bar{\sigma}}^o+\eth^2\bar{\sigma}^o)=0\\
&&\Psi^o_3=-\eth\dot{\bar{\sigma}}^o-\frac{\Lambda}{6}\bar{\Psi}^o_1+\frac{\Lambda}{3}\sigma^o\eth'\bar{\sigma}^o+\eth'U^o_1\\
&&\Psi^o_4=-\ddot{\bar{\sigma}}^o+\frac{\Lambda}{3}K\bar{\sigma}^o-\frac{\Lambda}{3}\eth'\eth\bar{\sigma}^o+\frac{2\Lambda^2}{9}\bar{\sigma}^o|\sigma^o|^2+\frac{\Lambda^2}{36}\bar{\Psi}^o_0\label{psi4ST}\\
&&\dot{\alpha}^o=\frac{\Lambda}{3}\bar{\alpha}^o\bar{\sigma}^o+\frac{\Lambda}{6}\eth\bar{\sigma}^o\\
&&(\dot{\xi}^\mu)^o=-\frac{\Lambda}{3}\sigma^o(\overline{\xi^\mu})^o\\
&&\textrm{Im}(\eth'(\xi^\mu)^o)=0\\
&&K=1+\frac{2\Lambda}{3}\int{\text{Re}(\eth^2\bar{\sigma}^o)du}\label{KST},
\end{eqnarray}
where $U^o_1=-\frac{1}{2}K-\frac{\Lambda}{2}|\sigma^o|^2$, and $K$ is the Gauss curvature of the compact 2-surface of constant $u$ on $\mathcal{I}$.

The dyad components of the Maxwell spinor are:
\begin{eqnarray}
\phi_0&=&\phi^o_0r^{-2}+\phi^1_0r^{-3}+O(r^{-4})\\
\phi_1&=&\phi^o_1r^{-2}-\eth'\phi^o_0r^{-3}+O(r^{-4})\\
\phi_2&=&\phi^o_2r^{-2}-\left(\eth'\phi^o_1-\frac{\Lambda}{6}\bar{\sigma}^o\phi^o_0\right)r^{-3}+O(r^{-4}),
\end{eqnarray}
with Maxwell equations:
\begin{eqnarray}
\dot{\phi}^o_0&=&\eth\phi^o_1+\sigma^o\phi^o_2+\frac{\Lambda}{6}\phi^1_0\label{max1ST}\\
\dot{\phi}^o_1&=&\eth\phi^o_2-\frac{\Lambda}{6}\eth'\phi^o_0.
\end{eqnarray}

Note that \emph{identical leading-order relations} in Eqs. (\ref{Biatwo}) and (\ref{BiatwoST}) are obtained (amongst several other pairs of equations --- one from each null tetrad), as these are independent of $r$, despite differing fall-offs of various quantities in these different null tetrads. This should be expected, of course, since physics is independent of the choice of null tetrads used to describe it. More significantly, the matching of the leading-order relations Eqs. (\ref{Biatwo}) and (\ref{BiatwoST}) echoes Bondi et al.'s original work \cite{Bondi62}, viz. the resulting mass-loss formula is an \emph{exact result} (instead of being some approximation), although this considers an asymptotic expansion over large distances from the source.

\section{Mass-loss formula for an isolated electro-gravitating system with cosmological constant}

The Bondi mass for asymptotically flat spacetimes is $\displaystyle M_B=-\frac{1}{A}\oint{(\Psi^o_2+\sigma^o\dot{\bar{\sigma}}^o)}d^2S$, where integration is over a compact 2-surface of constant $u$ on $\mathcal{I}$ and $A$ is the area. A generalisation to asymptotically de Sitter spacetimes is given in Ref. \cite{Vee2016}:
\begin{eqnarray}
M_{\Lambda}:=M_B+\frac{1}{A}\int{\left(\oint{\left(\Psi^o_2+\sigma^o\dot{\bar{\sigma}}^o\right)\frac{\partial}{\partial u}(d^2S)}\right)du}+\frac{\Lambda}{3A}\int{\left(\oint{K|\sigma^o|^2d^2S}\right)du},
\end{eqnarray}
which ensures that this mass $M_\Lambda$ strictly decreases whenever $\sigma^o$ is non-zero in the absence of incoming gravitational radiation, i.e. the mass of an isolated gravitating system strictly decreases due to energy carried away by gravitational waves. From the Bianchi identity Eq. (\ref{Biatwo}), we have:
\begin{eqnarray}
\frac{dM_{\Lambda}}{du}
&=&-\frac{1}{A}\oint{\left(|\dot{\sigma}^o|^2+k|\phi^o_2|^2+\frac{\Lambda}{3}|\eth'\sigma^o|^2+\frac{2\Lambda^2}{9}|\sigma^o|^4+\frac{\Lambda^2}{18}\textrm{Re}(\bar{\sigma}^o\Psi^o_0)-\frac{k\Lambda^2}{36}|\phi^o_0|^2\right)d^2S}.\label{0Bondimasslosspsi0ST}\nonumber\\
\end{eqnarray}
If there are only Maxwell fields but no gravitational radiation, i.e. $\sigma^o=0$, then
\begin{eqnarray}
\frac{dM_{\Lambda}}{du}
&=&-\frac{1}{A}\oint{\left(k|\phi^o_2|^2-\frac{k\Lambda^2}{36}|\phi^o_0|^2\right)d^2S},\label{0BondimasslossMaxonly}
\end{eqnarray}
where $M_\Lambda=M_B$. Hence, outgoing electromagnetic (EM) radiation $\phi^o_2$ causes the mass of the isolated system to decrease, with incoming EM radiation $\phi^o_0$ giving rise to an increase in mass \footnote{For gravitation, $\Psi^o_4$ represents outgoing (transverse) radiation whilst $\Psi^o_0$ represents incoming (transverse) radiation. The Coulomb term is $\Psi^o_2$ \cite{Ste}. For electromagnetism, $\phi^o_2$ represents outgoing radiation, whilst $\phi^o_0$ represents incoming radiation. The Coulomb term is $\phi^o_1$.}. Unlike the gravitational counterpart, there is no unusual Maxwell coupling with the cosmological constant other than $k\Lambda^2|\phi^o_0|^2/36$. In particular, the presence of Maxwell fields does not alter the structure of $\mathcal{I}$. Note that for gravitation, outgoing gravitational radiation can only carry energy away from the isolated source if the spacelike $\mathcal{I}$ is non-conformally flat \cite{Vee2016,ash1}. Such differing properties are perhaps manifestations of the linear Maxwell fields versus the non-linear Einstein fields. The only EM coupling to the cosmological constant is due to $\mathcal{I}$ being spacelike for $\Lambda>0$, such that incoming EM radiation $\phi^o_0$ beyond the isolated system's cosmological horizon would arrive at $\mathcal{I}$ and get picked up by the mass-loss formula. No incoming EM radiation corresponds to setting $\phi^o_0=0$. The Maxwell equation Eq. ({\ref{max1}}) does not constrain $\eth\phi^o_1$ and $\sigma^o\phi^o_2$, since there is a term involving $\Lambda\phi^1_0/6$ which now plays the role of $\dot{\phi^o_0}$ (had $\phi^o_0$ not been set to zero). This is analogous to ``no-incoming-gravitational-radiation'' being $\Psi^o_0=0$ \cite{Vee2016}. Apart from that, integrating Eq. (\ref{max2}) over a compact 2-surface of constant $u$ on $\mathcal{I}$ gives charge conservation.

\section{Concluding remarks}

In the spirit of Newman and Unti, we presented the asymptotic solutions to the Newman-Penrose equations for asymptotically de Sitter spacetimes containing Maxwell fields. One can directly check by plugging them into all 42 Newman-Penrose equations, to verify that they are all satisfied (up to the relevant orders). These extend a recent work which solved for the behaviour of asymptotically empty de Sitter spacetimes, where the asymptotic solutions here reduce to those in Ref. \cite{Vee2016} in the absence of the Maxwell fields. Also, the asymptotic solutions presented here reduce to those of the Einstein-Maxwell equations presented in Ref. \cite{Pen88}, for $\Lambda=0$. Those interested in the full details for the derivation of these asymptotic solutions by solving the 42 Newman-Penrose-Maxwell equations (in the same manner as was done in Ref. \cite{Vee2016} for empty spacetimes) may refer to Appendix B.

\begin{acknowledgments}
I wish to thank J\"{o}rg Frauendiener for some discussions leading to better clarity and presentation of the manuscript. V.-L. Saw is supported by the University of Otago Doctoral Scholarship.
\end{acknowledgments}

\newpage

\appendix

\section{Boost transformation}

Here is a list of transformations under $\vec{l}$-$\vec{n}$ rescaling (boost transformation). Consider a new null tetrad under the transformation $l^a\rightarrow f(r)l^a$, $\displaystyle n^a\rightarrow\frac{1}{f(r)}n^a$, $l_a\rightarrow f(r)l_a$, $\displaystyle n_a\rightarrow\frac{1}{f(r)}n_a$, where $f(r)$ is a well-behaved function of $r$. The vectors $\vec{m}$ and $\vec{\bar{m}}$ are unchanged.

The spin-coefficients are:
\begin{eqnarray}
\kappa&=&m^aDl_a=m^al^b\nabla_bl_a\rightarrow m^a(f(r)l^b)\nabla_b(f(r)l_a)=f(r)^2\kappa\\
\kappa'&=&\bar{m}^aD'n_a=\bar{m}^an^b\nabla_bn_a\rightarrow\bar{m}^a\left(\frac{1}{f(r)}n^b\right)\nabla_b\left(\frac{1}{f(r)}n_a\right)=\frac{1}{f(r)^2}\kappa'\\
\sigma&=&m^a\delta l_a\rightarrow m^a\delta(f(r)l_a)=f(r)\sigma\\
\sigma'&=&\bar{m}^a\delta'n_a\rightarrow\bar{m}^a\delta'\left(\frac{1}{f(r)}n_a\right)=\frac{1}{f(r)}\sigma'\\
\tau&=&m^aD'l_a=m^an^b\nabla_bl_a\rightarrow m^a\left(\frac{1}{f(r)}n^b\right)\nabla_b(f(r)l_a)=\tau\\
\tau'&=&\bar{m}^aDn_a=\bar{m}^al^b\nabla_bn_a\rightarrow\bar{m}^a(f(r)l^b)\nabla_b\left(\frac{1}{f(r)}n_a\right)=\tau'\\
\rho&=&m^a\delta'l_a\rightarrow m^a\delta'(f(r)l_a)=f(r)\rho\\
\rho'&=&\bar{m}^a\delta n_a\rightarrow\bar{m}^a\delta\left(\frac{1}{f(r)}n_a\right)=\frac{1}{f(r)}\rho'\\
\gamma&=&\frac{1}{2}(n^aD'l_a-\bar{m}^aD'm_a)=\frac{1}{2}(n^an^b\nabla_bl_a-\bar{m}^an^b\nabla_bm_a)\nonumber\\&\rightarrow&\frac{1}{2}\left(\left(\frac{1}{f(r)}n^a\right)\left(\frac{1}{f(r)n^b}\right)\nabla_b(f(r)l_a)-\bar{m}^a\left(\frac{1}{f(r)}n^b\right)\nabla_bm_a\right)=\frac{1}{f(r)}\gamma+\frac{D'(f(r))}{2f(r)^2}\nonumber\\&=&\frac{1}{f(r)}\gamma+\frac{f'(r)}{2f(r)^2}U\\
\gamma'&=&\frac{1}{2}(l^aDn_a-m^aD\bar{m}_a)=\frac{1}{2}(l^al^b\nabla_bn_a-m^al^b\nabla_b\bar{m}_a)\nonumber\\&\rightarrow&\frac{1}{2}\left((f(r)l^a)(f(r)l^b)\nabla_b\left(\frac{1}{f(r)}n_a\right)-m^a(f(r)l^b)\nabla_b\bar{m}_a\right)=f(r)\gamma'+\frac{1}{2}f(r)^2D\left(\frac{1}{f(r)}\right)\nonumber\\&=&f(r)\gamma'-\frac{1}{2}f'(r)\\
\alpha&=&\frac{1}{2}(n^a\delta'l_a-\bar{m}^a\delta'm_a)\rightarrow\frac{1}{2}\left(\left(\frac{1}{f(r)}n^a\right)\delta'(f(r)l_a)-\bar{m}^a\delta'm_a\right)=\alpha+\frac{1}{2f(r)}\delta'(f(r))\nonumber\\&=&\alpha+\frac{f'(r)}{2f(r)}\bar{\omega}\\
\alpha'&=&\frac{1}{2}(l^a\delta n_a-m^a\delta\bar{m}_a)\rightarrow\frac{1}{2}\left((f(r)l^a)\delta\left(\frac{1}{f(r)}n_a\right)-m^a\delta\bar{m}^a\right)=\alpha'+\frac{1}{2}f(r)\delta\left(\frac{1}{f(r)}\right)\nonumber\\&=&\alpha'-\frac{f'(r)}{2f(r)}\omega.
\end{eqnarray}

\newpage

The dyad components of the Weyl spinor are:
\begin{eqnarray}
\Psi_0&=&C_{abcd}l^am^bl^cm^d\rightarrow C_{abcd}(f(r)l^a)m^b(f(r)l^c)m^d=f(r)^2\Psi_0\\
\Psi_1&=&C_{abcd}l^am^bl^cn^d\rightarrow C_{abcd}(f(r)l^a)m^b(f(r)l^c)\left(\frac{1}{f(r)}n^d\right)=f(r)\Psi_1\\
\Psi_2&=&C_{abcd}l^am^b\bar{m}^cn^d\rightarrow C_{abcd}(f(r)l^a)m^b\bar{m}^c\left(\frac{1}{f(r)}n^d\right)=\Psi_2\\
\Psi_3&=&C_{abcd}l^an^b\bar{m}^cn^d\rightarrow C_{abcd}(f(r)l^a)\left(\frac{1}{f(r)}n^b\right)\bar{m}^c\left(\frac{1}{f(r)}n^d\right)=\frac{1}{f(r)}\Psi_3\\
\Psi_4&=&C_{abcd}\bar{m}^an^b\bar{m}^cn^d\rightarrow C_{abcd}\bar{m}^a\left(\frac{1}{f(r)}n^b\right)\bar{m}^c\left(\frac{1}{f(r)}n^d\right)=\frac{1}{f(r)^2}\Psi_4.
\end{eqnarray}

The dyad components of the Maxwell spinor are:
\begin{eqnarray}
\phi_0&=&F_{ab}l^am^b\rightarrow F_{ab}(f(r)l^a)m^b=f(r)\phi_0\\
\phi_1&=&\frac{1}{2}F_{ab}(l^an^b+\bar{m}^am^b)\rightarrow\frac{1}{2}F_{ab}\left((f(r)l^a)\left(\frac{1}{f(r)}n^b\right)+\bar{m}^am^b\right)=\phi_1\\
\phi_2&=&F_{ab}\bar{m}^an^b\rightarrow F_{ab}\bar{m}^a\left(\frac{1}{f(r)}n^b\right)=\frac{1}{f(r)}\phi_2.
\end{eqnarray}

For $f(r)=r^n$, where $n$ is an integer, the null tetrad vectors transform as:
\begin{eqnarray}
\vec{l}&\rightarrow&r^n\vec{l}\\
\vec{n}&\rightarrow&r^{-n}\vec{n}\\
\vec{m}&\rightarrow&\vec{m}\\
\vec{\bar{m}}&\rightarrow&\vec{\bar{m}}.
\end{eqnarray}

\newpage

The spin coefficients transform as:
\begin{eqnarray}
\kappa&\rightarrow&r^{2n}\kappa\\
\kappa'&\rightarrow&r^{-2n}\kappa'\\
\sigma&\rightarrow&r^n\sigma\\
\sigma'&\rightarrow&r^{-n}\sigma'\\
\tau&\rightarrow&\tau\\
\tau'&\rightarrow&\tau'\\
\rho&\rightarrow&r^n\rho\\
\rho'&\rightarrow&r^{-n}\rho\\
\gamma&\rightarrow&r^{-n}\gamma+\frac{1}{2}nr^{-n-1}U\\
\gamma'&\rightarrow&r^n\gamma'-\frac{1}{2}nr^{n-1}\\
\alpha&\rightarrow&\alpha+\frac{1}{2}nr^{-1}\bar{\omega}\\
\alpha'&\rightarrow&\alpha'-\frac{1}{2}nr^{-1}\omega.
\end{eqnarray}

The dyad components of the Weyl spinor transform as:
\begin{eqnarray}
\Psi_0&\rightarrow&r^{2n}\Psi_0\\
\Psi_1&\rightarrow&r^n\Psi_1\\
\Psi_2&\rightarrow&\Psi_2\\
\Psi_3&\rightarrow&r^{-n}\Psi_3\\
\Psi_4&\rightarrow&r^{-2n}\Psi_4.
\end{eqnarray}

The dyad components of the Maxwell spinor transform as:
\begin{eqnarray}
\phi_0&\rightarrow&r^n\phi_0\\
\phi_1&\rightarrow&\phi_1\\
\phi_2&\rightarrow&r^{-n}\phi_2.
\end{eqnarray}

So we see that the fall-offs can be altered, depending on $n$ in $r^n\vec{l}$ and $r^{-n}\vec{n}$. There are some quantities which remain invariant, viz. $\Psi_2, \phi_1, \tau, \tau'$; with $\alpha, \alpha'$ getting additional terms from order $r^{-2}$  onwards.

For $n=1$, the resulting null tetrad is that of Szabados-Tod \cite{Szabados} which corresponds to the symmetric scaling of the conformal factor over the spinor dyad $o^A$ and $\iota^A$.

\section{Details on solving the 42 equations of the Newman-Penrose formalism (metric equations, spin coefficient equations, Bianchi identities, Maxwell equations), order-by-order away from $\mathcal{I}$}

Asymptotically de Sitter spacetimes have a null tetrad of the form \cite{Vee2016}:
\begin{eqnarray}
\vec{l}&=&\vec{\partial}_r\\
\vec{n}&=&\vec{\partial}_u+U\vec{\partial}_r+X^\theta\vec{\partial}_\theta+X^\phi\vec{\partial}_\phi\\
\vec{m}&=&\omega\vec{\partial}_r+\xi^\theta\vec{\partial}_\theta+\xi^\phi\vec{\partial}_\phi\\
\vec{\bar{m}}&=&\bar{\omega}\vec{\partial}_r+\overline{\xi^\theta}\vec{\partial}_\theta+\overline{\xi^\phi}\vec{\partial}_\phi,
\end{eqnarray}
where
\begin{eqnarray}
U&=&\frac{\Lambda}{6}r^2+U^o_0+O(r^{-1})\\
X^\mu&=&(X^\mu)^o_1r^{-1}+O(r^{-2})\\
\omega&=&\omega^o_1r^{-1}+O(r^{-2})\\
\xi^\mu&=&(\xi^\mu)^o_1r^{-1}+O(r^{-2}).
\end{eqnarray}
In writing this, we adopt the notation that a function $f(u,r,\theta,\phi)$ can be expanded as a series $\displaystyle f(u,r,\theta,\phi)=\sum_i^N{f^o_{i}(u,\theta,\phi)r^{-i}}$ in $r$ for sufficiently large $N$, and the superscript $^o$ means that the coefficients $f^o_i$ are independent of $r$.

The directional derivatives are
\begin{eqnarray}
D&=&l^a\nabla_a=\frac{\partial}{\partial r}\\
D'&=&n^a\nabla_a=\frac{\partial}{\partial u}+U\frac{\partial}{\partial r}+X^\theta\frac{\partial}{\partial\theta}+X^\phi\frac{\partial}{\partial\phi}\\
\delta&=&m^a\nabla_a=\omega\frac{\partial}{\partial r}+\xi^\theta\frac{\partial}{\partial\theta}+\xi^\phi\frac{\partial}{\partial\phi}\\
\delta'&=&\bar{m}^a\nabla_a=\bar{\omega}\frac{\partial}{\partial r}+\overline{\xi^\theta}\frac{\partial}{\partial\theta}+\overline{\xi^\phi}\frac{\partial}{\partial\phi}.
\end{eqnarray}

In the Newman-Penrose formalism, the Christoffel symbols (connection coefficients) are re-expressed in terms of twelve complex spin coefficients, defined as \cite{Pen87,don}
\begin{eqnarray}
\kappa&=&m^aDl_a\\
\kappa'&=&\bar{m}^aD'n_a\\
\sigma&=&m^a\delta l_a\\
\sigma'&=&\bar{m}^a\delta'n_a\\
\tau&=&m^aD'l_a\\
\tau'&=&\bar{m}^aDn_a\\
\rho&=&m^a\delta'l_a\\
\rho'&=&\bar{m}^a\delta n_a\\
\gamma&=&\frac{1}{2}(n^aD'l_a-\bar{m}^aD'm_a)\\
\gamma'&=&\frac{1}{2}(l^aDn_a-m^aD\bar{m}_a)\\
\alpha&=&\frac{1}{2}(n^a\delta'l_a-\bar{m}^a\delta'm_a)\\
\alpha'&=&\frac{1}{2}(l^a\delta n_a-m^a\delta\bar{m}_a).
\end{eqnarray}

The spin coefficients for asymptotically de Sitter spacetimes have the form
\begin{eqnarray}
\kappa&=&0, \gamma'=0, \tau'=0,\\
\kappa'&=&\kappa'^o_0+\kappa'^o_1r^{-1}+O(r^{-2})\\
\sigma&=&\sigma^o_1r^{-1}+O(r^{-2})\\
\sigma'&=&\sigma'^o_0+\sigma'^o_1r^{-1}+O(r^{-2})\\
\tau&=&\tau^o_1r^{-1}+O(r^{-2})\\
\gamma&=&\gamma^o_{-1}r+\gamma^o_2r^{-2}+O(r^{-3}), \gamma^o_{-1}\neq0\\
\rho&=&\rho^o_1r^{-1}+\rho^o_3r^{-3}+O(r^{-4}), \rho^o_1\neq0\\
\rho'&=&\rho'^o_{-1}r+\rho'^o_1r^{-1}+O(r^{-2}), \rho'^o_{-1}\neq0\\
\alpha&=&\alpha^o_1r^{-1}+O(r^{-2}), \alpha^o_1\neq0\\
\alpha'&=&\alpha'^o_1r^{-1}+O(r^{-2}), \alpha'^o_1\neq0.
\end{eqnarray}

The Einstein field equations relate part of the spacetime curvature (viz. the Einstein tensor) to the stress-energy tensor. The remaining part of the (Riemann) curvature is contained in the Weyl tensor. In the Newman-Penrose formalism, it is convenient to introduce the Ricci spinor and the Weyl spinor instead, with the dyad components $\Phi_{ab}$ and $\Psi_a$ defined as
\begin{eqnarray}
\Phi_{00}&=&-\frac{1}{2}R_{ab}l^al^b\\
\Phi_{01}&=&-\frac{1}{2}R_{ab}l^am^b\\
\Phi_{02}&=&-\frac{1}{2}R_{ab}m^am^b\\
\Phi_{10}&=&-\frac{1}{2}R_{ab}l^a\bar{m}^b\\
\Phi_{11}&=&-\frac{1}{2}R_{ab}l^an^b+\frac{1}{8}R\\
\Phi_{12}&=&-\frac{1}{2}R_{ab}m^an^b\\
\Phi_{20}&=&-\frac{1}{2}R_{ab}\bar{m}^a\bar{m}^b\\
\Phi_{21}&=&-\frac{1}{2}R_{ab}\bar{m}^an^b\\
\Phi_{22}&=&-\frac{1}{2}R_{ab}n^an^b,
\end{eqnarray}
\begin{eqnarray}
\Psi_0&=&C_{abcd}l^am^bl^cm^d\label{Weylstart}\\
\Psi_1&=&C_{abcd}l^am^bl^cn^d\\
\Psi_2&=&C_{abcd}l^am^b\bar{m}^cn^d\\
\Psi_3&=&C_{abcd}l^an^b\bar{m}^cn^d\\
\Psi_4&=&C_{abcd}\bar{m}^an^b\bar{m}^cn^d\label{Weylend},
\end{eqnarray}
where $R_{ab}$ is the Ricci tensor, $R$ is the Ricci scalar, $\Lambda$ is the cosmological constant, and $C_{abcd}$ is the Weyl tensor. Since the Maxwell stress-energy tensor is traceless, the Einstein-Maxwell field equations with cosmological constant gives $\displaystyle\Lambda=\frac{1}{4}R$.

The Maxwell spinor has dyad components defined as
\begin{eqnarray}
\phi_0&=&F_{ab}l^am^b\\
\phi_1&=&\frac{1}{2}F_{ab}(l^an^b+\bar{m}^am^b)\\
\phi_2&=&F_{ab}\bar{m}^an^b,
\end{eqnarray}
where $F_{ab}$ is the electromagnetic field tensor.

In an electro-$\Lambda$ spacetime, the Ricci spinor is proportional to the Maxwell spinor \cite{newpen62}, so that
\begin{eqnarray}\label{coupleMaxwell}
\Phi_{ab}=k\phi_a\bar{\phi}_b,
\end{eqnarray}
where $k=2G/c^4$.

Following the case for asymptotically flat spacetimes \cite{newpen62,newunti62}, we take $\Psi_0$ to have the fall-off of order $r^{-5}$, i.e. $\Psi_0=(\Psi_0)^o_5r^{-5}+O(r^{-6})$, and $\phi_0$ to have the fall-off of order $r^{-3}$, i.e. $\phi_0=(\phi_0)^o_3r^{-3}+O(r^{-4})$. We denote $\Psi^o_0:=(\Psi_0)^o_5$, $\Psi^1_0:=(\Psi_0)^o_6$, $\Psi^o_1:=(\Psi_1)^o_4$, $\Psi^o_2:=(\Psi_2)^o_3$, $\Psi^o_3:=(\Psi_3)^o_2$, $\Psi^o_4:=(\Psi_4)^o_1$, $\phi^o_0:=(\phi_0)^o_3$, $\phi^1_0:=(\phi_0)^o_4$, $\phi^o_1:=(\phi_1)^o_2$, $\phi^o_2:=(\phi_2)^o_1$. Furthermore, we adopt the following notation:
\begin{eqnarray}
\delta^o:=(\xi^\theta)^o\frac{\partial}{\partial\theta}+(\xi^\phi)^o\frac{\partial}{\partial\phi}, \delta'^o:=(\overline{\xi^\theta})^o\frac{\partial}{\partial\theta}+(\overline{\xi^\phi})^o\frac{\partial}{\partial\phi},
\end{eqnarray}
where $(\xi^\theta)^o:=(\xi^\theta)^o_1$, $(\xi^\phi)^o:=(\xi^\phi)^o_1$ and thus define the $\eth$ and $\eth'$ operators acting on a scalar $\eta$ with spin-weight $s$ to be \cite{Pen87}:
\begin{eqnarray}
\eth\eta:=(\delta^o+2s\bar{\alpha}^o)\eta, \eth'\eta:=(\delta'^o-2s\alpha^o)\eta,
\end{eqnarray}
where $\alpha^o:=\alpha^o_1$.

The metric equations are
\begin{eqnarray}
{[}\delta,D]r &:& (11)\ D\omega=\rho\omega+\sigma\bar{\omega}+\alpha'-\bar{\alpha}\\
{[}D',D]r &:& (16)\ DU=2\textrm{Re}(\bar{\tau}\omega-\gamma)\\
{[}\delta,D]\theta &:& (18)\ D\xi^\theta=\rho\xi^\theta+\sigma\overline{\xi^\theta}\\
{[}\delta,D]\phi &:& (19)\ D\xi^\phi=\rho\xi^\phi+\sigma\overline{\xi^\phi}\\
{[}D',D]\theta &:& (20)\ DX^\theta=2\textrm{Re}(\bar{\tau}\xi^\theta)\\
{[}D',D]\phi &:& (21)\ DX^\phi=2\textrm{Re}(\bar{\tau}\xi^\phi)\\
{[}\delta,D']r &:& (22)\ D'\omega-\delta U=(\rho'+2i\textrm{Im}(\gamma))\omega+\bar{\sigma}'\bar{\omega}-\bar{\kappa}'\\
{[}\delta',\delta]r &:& (23)\ \textrm{Im}(\delta'\omega)=\textrm{Im}(\rho'+(\alpha+\bar{\alpha}')\omega)\\
{[}\delta,D']\theta &:& (28)\ D'\xi^\theta-\delta X^\theta=(\rho'+2i\textrm{Im}(\gamma))\xi^\theta+\bar{\sigma}'\overline{\xi^\theta}\\
{[}\delta,D']\phi &:& (29)\ D'\xi^\phi-\delta X^\phi=(\rho'+2i\textrm{Im}(\gamma))\xi^\phi+\bar{\sigma}'\overline{\xi^\phi}\\
{[}\delta',\delta]\theta &:& (34)\ \textrm{Im}(\delta'\xi^\theta)=\textrm{Im}((\alpha+\bar{\alpha}')\xi^\theta)\\
{[}\delta',\delta]\phi &:& (35)\ \textrm{Im}(\delta'\xi^\phi)=\textrm{Im}((\alpha+\bar{\alpha}')\xi^\phi).
\end{eqnarray}

\newpage

The spin coefficient equations are
\begin{eqnarray}
(1)\ D\sigma'&=&\sigma'\rho+\rho'\bar{\sigma}-k\phi_2\bar{\phi}_0\\
(2)\ D\rho&=&\rho^2+\sigma\bar{\sigma}+k\phi_0\bar{\phi}_0\\
(3)\ D\sigma&=&2\rho\sigma+\Psi_0\\
(10)\ D\rho'&=&\rho'\rho+\sigma'\sigma-\Psi_2-\frac{\Lambda}{3}\\
(12)\ D\alpha&=&\alpha\rho-\alpha'\bar{\sigma}+k\phi_1\bar{\phi}_0\\
(13)\ D\alpha'&=&\alpha'\rho-\alpha\sigma-\Psi_1\\
(14)\ D\tau&=&\tau\rho+\bar{\tau}\sigma+\Psi_1+k\phi_0\bar{\phi}_1\ (\tau=\bar{\alpha}-\alpha')\\
(15)\ D\gamma&=&\tau\alpha-\bar{\tau}\alpha'+\Psi_2-\frac{\Lambda}{6}+k\phi_1\bar{\phi}_1\\
(17)\ D\kappa'&=&\tau\sigma'+\bar{\tau}\rho'-\Psi_3-k\phi_2\bar{\phi}_1\\
(24)\ D'\sigma-\delta\tau&=&\rho\bar{\sigma}'+\sigma(\rho'+2\gamma+2i\textrm{Im}(\gamma))+2\alpha'\tau-k\phi_0\bar{\phi}_2\\
(25)\ D'\rho-\delta'\tau&=&\sigma\sigma'+\rho(\bar{\rho}'+2\textrm{Re}(\gamma))-2\alpha\tau-\Psi_2-\frac{\Lambda}{3}\\
(26)\ D'\rho'-\delta\kappa'&=&\sigma'\bar{\sigma}'+\rho'(\rho'-2\textrm{Re}(\gamma))-2\alpha'\kappa'+k\phi_2\bar{\phi}_2\\
(27)\ D'\sigma'-\delta'\kappa'&=&\sigma'(2\textrm{Re}(\rho')-2\gamma-2i\textrm{Im}(\gamma))+2\alpha\kappa'+\Psi_4\\
(30)\ D'\alpha-\delta'\gamma&=&\alpha(\bar{\rho}'-2i\textrm{Im}(\gamma))+(\bar{\alpha}-2\alpha')\sigma'-\rho\kappa'-\Psi_3\\
(31)\ D'\alpha'+\delta\gamma&=&\alpha'(\rho'+2i\textrm{Im}(\gamma))-\alpha\bar{\sigma}'+\sigma\kappa'-\tau\rho'+k\phi_1\bar{\phi}_2\\
(32)\ \delta\rho-\delta'\sigma&=&(\bar{\alpha}-\alpha')\rho-(3\alpha+\bar{\alpha}')\sigma-\Psi_1+k\phi_0\bar{\phi}_1\\
(33)\ \delta'\rho'-\delta\sigma'&=&(\bar{\alpha}'-\alpha)\rho'-(3\alpha'+\bar{\alpha})\sigma'-\Psi_3+k\phi_2\bar{\phi}_1\\
(36)\ \delta\alpha+\delta'\alpha'&=&\alpha\bar{\alpha}+\alpha'\bar{\alpha}'+2\alpha\alpha'-\rho\rho'+\sigma\sigma'-\Psi_2+\frac{\Lambda}{6}+k\phi_1\bar{\phi}_1.
\end{eqnarray}

\newpage

The Bianchi identities are
\begin{eqnarray}
(6)\ (D-4\rho)\Psi_1-(\delta'-4\alpha)\Psi_0&=&-k\bar{\phi}_0(\delta+2\alpha')\phi_0+k\bar{\phi}_1D\phi_0\nonumber\\&&-2k\sigma\phi_1\bar{\phi}_0\\
(7)\ (D-3\rho)\Psi_2-(\delta'-2\alpha)\Psi_1-\sigma'\Psi_0&=&-k\bar{\phi}_0(D'-2\gamma)\phi_0+k\bar{\phi}_1(\delta'-2\alpha)\phi_0\nonumber\\&&+2k\phi_1(\rho\bar{\phi}_1-\tau\bar{\phi}_0)\\
(8)\ (D-2\rho)\Psi_3-\delta'\Psi_2-2\sigma'\Psi_1&=&-k\bar{\phi}_0(\delta-2\alpha')\phi_2+k\bar{\phi}_1D\phi_2\nonumber\\&&-2k\rho'\phi_1\bar{\phi}_0\\
(9)\ (D-\rho)\Psi_4-(\delta'+2\alpha)\Psi_3-3\sigma'\Psi_2&=&-k\bar{\phi}_0(D'+2\gamma)\phi_2+k\bar{\phi}_1(\delta'+2\alpha)\phi_2\nonumber\\&&+2k\phi_1(\sigma'\bar{\phi}_1-\kappa'\bar{\phi}_0)\\
(39)\ (D'-4\gamma-\rho')\Psi_0-(\delta-4\tau+2\alpha')\Psi_1-3\sigma\Psi_2&=&k\bar{\phi}_1(\delta+2\alpha')\phi_0-k\bar{\phi}_2D\phi_0\nonumber\\&&+2k\sigma\phi_1\bar{\phi}_1\\
(40)\ \kappa'\Psi_0+(D'-2\gamma-2\rho')\Psi_1-(\delta-3\tau)\Psi_2-2\sigma\Psi_3&=&k\bar{\phi}_1(D'-2\gamma)\phi_0-k\bar{\phi}_2(\delta'-2\alpha)\phi_0\nonumber\\&&+2k\phi_1(\tau\bar{\phi}_1-\rho\bar{\phi}_2)\\
(41)\ 2\kappa'\Psi_1+(D'-3\rho')\Psi_2-(\delta-2\tau-2\alpha')\Psi_3-\sigma\Psi_4&=&k\bar{\phi}_1(\delta-2\alpha')\phi_2-k\bar{\phi}_2D\phi_2\nonumber\\&&+2k\rho'\phi_1\bar{\phi}_1\\
(42)\ 3\kappa'\Psi_2+(D'+2\gamma-4\rho')\Psi_3-(\delta-\tau-4\alpha')\Psi_4&=&k\bar{\phi}_1(D'+2\gamma)\phi_2-k\bar{\phi}_2(\delta'+2\alpha)\phi_2\nonumber\\&&+2k\phi_1(\kappa'\bar{\phi}_1-\sigma'\bar{\phi}_2).
\end{eqnarray}

Maxwell equations are
\begin{eqnarray}
(4)\ (D-2\rho)\phi_1&=&(\delta'-2\alpha)\phi_0\\
(5)\ (D-\rho)\phi_2&=&\delta'\phi_1+\sigma'\phi_0\\
(37)\ (D'-2\gamma-\rho')\phi_0&=&(\delta-2\tau)\phi_1+\sigma\phi_2\\
(38)\ (D'-2\rho')\phi_1&=&(\delta-\tau-2\alpha')\phi_2-\kappa'\phi_0.
\end{eqnarray}

With the setup in place, the 42 equations of the Newman-Penrose formalism comprising the metric equations, spin coefficient equations, Bianchi identities, and Maxwell equations can be solved asymptotically away from $\mathcal{I}$. These are solved for various orders beginning from the highest non-trivial power of $r$, in the following sequence as indicated by the number in front of the equation. Listed below are the results when solving them order-by-order. We begin by solving the radial equations having the $D=\partial/\partial r$ derivative. Well, ``$0=0$'' means that there is no information or no \emph{new} information at this order, i.e. the equation is identically satisfied at this order.

(1) $D\sigma'=\sigma'\rho+\rho'\bar{\sigma}-k\phi_2\bar{\phi}_0$.
\begin{eqnarray}
1&:&\sigma^o_1=0\\
r^{-1}&:&\sigma'^o_0=-\frac{\rho'^o_{-1}}{\rho^o_1}\bar{\sigma}^o.\\
&\ &\textrm{So }\sigma'^o_0=-\frac{\Lambda}{6}\bar{\sigma}^o,\textrm{ from }\rho^o_1=-1\textrm{ in (2), }\rho'^o_{-1}=-\frac{\Lambda}{6}\textrm{ in (10).}\label{sigmaprime}
\end{eqnarray}

(2) $D\rho=\rho^2+\sigma\bar{\sigma}+k\phi_0\bar{\phi}_0$.

(3) $D\sigma=2\rho\sigma+\Psi_0$.
\begin{eqnarray}
r^{-2}&:&
\begin{cases}
\rho^o_1(\rho^o_1+1)=0,\textrm{ so }\rho^o_1=-1, \textrm{ since }\rho^o_1\neq0\\
0=0
\end{cases}\\
r^{-3}&:&
\begin{cases}
0=0\\
0=0
\end{cases}\\
r^{-4}&:&
\begin{cases}
\rho^o_3=-|\sigma^o|^2\\
\sigma^o_3=0
\end{cases}\\
r^{-5}&:&
\begin{cases}
\rho^o_4=0\\
\sigma^o_4=\sigma^o|\sigma^o|^2-\frac{1}{2}\Psi^o_0
\end{cases}\\
r^{-6}&:&
\begin{cases}
\rho^o_5=-|\sigma^o|^4+\frac{1}{3}\textrm{Re}(\bar{\sigma}^o\Psi^o_0)-\frac{1}{3}k|\phi^o_0|^2\\
\sigma^o_5=-\frac{1}{3}\Psi^1_0.
\end{cases}
\end{eqnarray}

(4) $(D-2\rho)\phi_1=(\delta'-2\alpha)\phi_0$.
\begin{eqnarray}
r^{-2}&:&(\phi_1)^o_1=0\\
r^{-3}&:&0=0,\textrm{ i.e. }\phi_1=\phi^o_1r^{-2}+O(r^{-3})\\
r^{-4}&:&(\phi_1)^o_3=-\eth'\phi^o_0\\
\end{eqnarray}

(5) $(D-\rho)\phi_2=\delta'\phi_1+\sigma'\phi_0$
\begin{eqnarray}
r^{-2}&:&0=0,\textrm{ i.e. }\phi_2=\phi^o_2r^{-1}+O(r^{-2})\\
r^{-3}&:&(\phi_2)^o_2=-\eth'\phi^o_1-\sigma'^o_0\phi^o_0\\
\end{eqnarray}

(6) $(D-4\rho)\Psi_1-(\delta'-4\alpha)\Psi_0=-k\bar{\phi}_0(\delta+2\alpha')\phi_0+k\bar{\phi}_1D\phi_0-2k\sigma\phi_1\bar{\phi}_0$.
\begin{eqnarray}
r^{-2}&:&(\Psi_1)^o_1=0\\
r^{-3}&:&(\Psi_1)^o_2=0\\
r^{-4}&:&(\Psi_1)^o_3=0\\
r^{-5}&:&0=0,\textrm{ i.e. }\Psi_1=\Psi^o_1r^{-4}+O(r^{-5})\\
r^{-6}&:&(\Psi_1)^o_5=-\eth'\Psi^o_0+3k\phi^o_0\bar{\phi}^o_1.
\end{eqnarray}

(7) $(D-3\rho)\Psi_2-(\delta'-2\alpha)\Psi_1-\sigma'\Psi_0=-k\bar{\phi}_0(D'-2\gamma)\phi_0+k\bar{\phi}_1(\delta'-2\alpha)\phi_0+2k\phi_1(\rho\bar{\phi}_1-\tau\bar{\phi}_0)$.
\begin{eqnarray}
r^{-2}&:& (\Psi_2)^o_1=0\\
r^{-3}&:&(\Psi_2)^o_2=0\\
r^{-4}&:&0=0,\textrm{ i.e. }\Psi_2=\Psi^o_2r^{-3}+O(r^{-4})\\
r^{-5}&:&(\Psi_2)^o_4=-\eth'\Psi^o_1-\sigma'^o_0\Psi^o_0+2k|\phi^o_1|^2-k\left(\frac{\Lambda}{2}+2\gamma^o_{-1}\right)|\phi^o_0|^2.
\end{eqnarray}

(8) $(D-2\rho)\Psi_3-\delta'\Psi_2-2\sigma'\Psi_1=-k\bar{\phi}_0(\delta-2\alpha')\phi_2+k\bar{\phi}_1D\phi_2-2k\rho'\phi_1\bar{\phi}_0$.
\begin{eqnarray}
r^{-2}&:& (\Psi_3)^o_1=0\\
r^{-3}&:&0=0,\textrm{ i.e. }\Psi_3=\Psi^o_3r^{-2}+O(r^{-3})\\
r^{-4}&:&(\Psi_3)^o_3=-\eth'\Psi^o_2-2\sigma'^o_0\Psi^o_1+k\phi^o_2\bar{\phi}^o_1+2k\rho'^o_{-1}\phi^o_1\bar{\phi}^o_0.
\end{eqnarray}

(9) $(D-\rho)\Psi_4-(\delta'+2\alpha)\Psi_3-3\sigma'\Psi_2=-k\bar{\phi}_0(D'+2\gamma)\phi_2+k\bar{\phi}_1(\delta'+2\alpha)\phi_2+2k\phi_1(\sigma'\bar{\phi}_1-\kappa'\bar{\phi}_0)$.
\begin{eqnarray}
r^{-2}&:&0=0,\textrm{ i.e. }\Psi_4=\Psi^o_4r^{-1}+O(r^{-2})\\
r^{-3}&:&(\Psi_4)^o_2=-\eth'\Psi^o_3-3\sigma'^o_0\Psi^o_2-k\left(\frac{\Lambda}{6}-2\gamma^o_{-1}\right)\phi^o_2\bar{\phi}^o_0.
\end{eqnarray}
Ergo, asymptotically de Sitter spacetimes with Maxwell fields have the peeling property, arising from the fall-offs of $\Psi_0$ being $O(r^{-5})$ and $\phi_0$ being $O(r^{-3})$, such that
\begin{eqnarray}
\Psi_n=O(r^{n-5}), \phi_m=O(r^{m-3}),
\end{eqnarray}
where $n=0,1,2,3,4$, $m=0,1,2$.

(10) $\displaystyle D\rho'=\rho'\rho+\sigma'\sigma-\Psi_2-\frac{\Lambda}{3}$.
\begin{eqnarray}
1&:&\rho'^o_{-1}=-\frac{\Lambda}{6}\\
r^{-1}&:&0=0\\
r^{-2}&:&0=0\\
r^{-3}&:&\rho'^o_2=\Psi^o_2-\sigma^o\sigma'^o_1\\
r^{-4}&:&\rho'^o_3=-\frac{1}{2}(\rho^o_3\rho'^o_1+\rho^o_5\rho'^o_{-1}+\sigma'^o_0\sigma^o_4+\sigma'^o_2\sigma^o-(\Psi_2)^o_4).
\end{eqnarray}

(11) $D\omega=\rho\omega+\sigma\bar{\omega}+\alpha'-\bar{\alpha}$.
\begin{eqnarray}
r^{-1}&:&\alpha'^o_1=\bar{\alpha}^o\\
r^{-2}&:&\alpha'^o_2=\bar{\alpha}^o_2\\
r^{-3}&:&\omega^o_2=\bar{\alpha}^o_3-\alpha'^o_3-\sigma^o\bar{\omega}^o_1=-\frac{1}{2}\Psi^o_1-\sigma^o\bar{\omega}^o_1,\textrm{ from }(12)\textrm{ and }(13).
\end{eqnarray}

(12) $D\alpha=\alpha\rho-\alpha'\bar{\sigma}+k\phi_1\bar{\phi}_0$.

(13) $D\alpha'=\alpha'\rho-\alpha\sigma-\Psi_1$.
\begin{eqnarray}
r^{-2}&:&
\begin{cases}
0=0\\
0=0
\end{cases}\\
r^{-3}&:&
\begin{cases}
\alpha^o_2=\bar{\alpha}^o\bar{\sigma}^o\\
\alpha'^o_2=\alpha^o\sigma^o
\end{cases}\\
r^{-4}&:&
\begin{cases}
\alpha^o_3=\alpha^o|\sigma^o|^2\\
\alpha'^o_3=\bar{\alpha}^o|\sigma^o|^2+\frac{1}{2}\Psi^o_1.
\end{cases}
\end{eqnarray}

(14) $\tau=\bar{\alpha}-\alpha'$.
\begin{eqnarray}
r^{-1}&:&\tau^o_1=0\\
r^{-2}&:&\tau^o_2=0\\
r^{-3}&:&\tau^o_3=-\frac{1}{2}\Psi^o_1.
\end{eqnarray}

(15) $\displaystyle D\gamma=\tau\alpha-\bar{\tau}\alpha'+\Psi_2-\frac{\Lambda}{6}+k\phi_1\bar{\phi}_1$.
\begin{eqnarray}
1&:&\gamma^o_{-1}=-\frac{\Lambda}{6}\\
r^{-1}&:&0=0\\
r^{-2}&:&0=0\\
r^{-3}&:&\gamma^o_2=-\frac{1}{2}\Psi^o_2.
\end{eqnarray}

(16) $DU=2\textrm{Re}(\bar{\tau}\omega-\gamma)$.
\begin{eqnarray}
r&:&0=0\\
1&:&0=0\\
r^{-1}&:&0=0\\
r^{-2}&:&U^o_1=2\textrm{Re}(\gamma^o_2)=-\textrm{Re}(\Psi^o_2)\\
r^{-3}&:&U^o_2=\textrm{Re}(\gamma^o_3).
\end{eqnarray}

(17) $D\kappa'=\tau\sigma'+\bar{\tau}\rho'-\Psi_3-k\phi_2\bar{\phi}_1$.
\begin{eqnarray}
r^{-2}&:&\kappa'^o_1=\Psi^o_3-\frac{\Lambda}{12}\bar{\Psi}^o_1\\
r^{-3}&:&2\kappa'^o_2=(\Psi_3)^o_3-\sigma'^o_0\tau^o_3-\bar{\tau}^o_4\rho'^o_{-1}+k\phi^o_2\bar{\phi}^o_1.
\end{eqnarray}

(18) $D\xi^\theta=\rho\xi^\theta+\sigma\overline{\xi^\theta}$.

(19) $D\xi^\phi=\rho\xi^\phi+\sigma\overline{\xi^\phi}$.
\begin{eqnarray}
r^{-2}&:&
\begin{cases}
0=0\\
0=0
\end{cases}\\
r^{-3}&:&
\begin{cases}
(\xi^\theta)^o_2=-\sigma^o(\overline{\xi^\theta})^o\\
(\xi^\phi)^o_2=-\sigma^o(\overline{\xi^\phi})^o
\end{cases}\\
r^{-4}&:&
\begin{cases}
(\xi^\theta)^o_3=|\sigma^o|^2(\xi^\theta)^o\\
(\xi^\phi)^o_3=|\sigma^o|^2(\xi^\phi)^o
\end{cases}\\
r^{-5}&:&
\begin{cases}
(\xi^\theta)^o_4=\left(\frac{1}{6}\Psi^o_0-\sigma^o|\sigma^o|^2\right)(\overline{\xi^\theta})^o\\
(\xi^\phi)^o_4=\left(\frac{1}{6}\Psi^o_0-\sigma^o|\sigma^o|^2\right)(\overline{\xi^\phi})^o.
\end{cases}
\end{eqnarray}

\newpage

(20) $DX^\theta=2\textrm{Re}(\bar{\tau}\xi^\theta)$.

(21) $DX^\phi=2\textrm{Re}(\bar{\tau}\xi^\phi)$.
\begin{eqnarray}
r^{-2}&:&
\begin{cases}
(X^\theta)^o_1=0\\
(X^\phi)^o_1=0
\end{cases}\\
r^{-3}&:&
\begin{cases}
(X^\theta)^o_2=0\\
(X^\phi)^o_2=0
\end{cases}\\
r^{-4}&:&
\begin{cases}
(X^\theta)^o_3=\frac{1}{3}\textrm{Re}(\bar{\Psi}^o_1(\xi^\theta)^o)\\
(X^\phi)^o_3=\frac{1}{3}\textrm{Re}(\bar{\Psi}^o_1(\xi^\phi)^o).
\end{cases}
\end{eqnarray}

This concludes solving all 21 radial equations involving the $D=\partial/\partial r$ derivative. Next, we deal with the other 21 equations.

(22) $D'\omega-\delta U=(\rho'+2i\textrm{Im}(\gamma))\omega+\bar{\sigma}'\bar{\omega}-\bar{\kappa}'$.
\begin{eqnarray}
1&:&\kappa'^o_0=\frac{\Lambda}{3}\bar{\omega}^o_1\label{kappaprime}\\
r^{-1}&:&\kappa'^o_1=-\dot{\bar{\omega}}^o_1-\frac{\Lambda}{4}\bar{\Psi}^o_1+\eth'U^o_0-\frac{2\Lambda}{3}\bar{\sigma}^o\omega^o_1.
\end{eqnarray}
Incidentally, (17) and (22) imply that
\begin{eqnarray}
\Psi^o_3=-\dot{\bar{\omega}}^o_1-\frac{\Lambda}{6}\bar{\Psi}^o_1+\eth'U^o_0-\frac{2\Lambda}{3}\bar{\sigma}^o\omega^o_1.\label{Psithree}
\end{eqnarray}

(23) $\textrm{Im}(\delta'\omega)=\textrm{Im}(\rho'+(\alpha+\bar{\alpha}')\omega)$.
\begin{eqnarray}
r^{-1}&:&\textrm{Im}(\rho'^o_1)=0\\
r^{-2}&:&\textrm{Im}(\rho'^o_2)=\textrm{Im}(\eth'\omega^o_1).
\end{eqnarray}

\newpage

(24) $D'\sigma-\delta\tau=\rho\bar{\sigma}'+\sigma(\rho'+2\gamma+2i\textrm{Im}(\gamma))+2\alpha'\tau-k\phi_0\bar{\phi}_2$.

(25) $\displaystyle D'\rho-\delta'\tau=\sigma\sigma'+\rho(\bar{\rho}'+2\textrm{Re}(\gamma))-2\alpha\tau-\Psi_2-\frac{\Lambda}{3}$.
\begin{eqnarray}
1&:&
\begin{cases}0=0\\
0=0
\end{cases}\\
r^{-1}&:&
\begin{cases}0=0\\
0=0
\end{cases}\\
r^{-2}&:&
\begin{cases}\sigma'^o_1=-\dot{\bar{\sigma}}^o\\
U^o_0=-\rho'^o_1-\frac{\Lambda}{6}|\sigma^o|^2
\end{cases}\\
r^{-3}&:&
\begin{cases}\sigma'^o_2=-\bar{\sigma}^o\rho'^o_1-\frac{\Lambda}{12}\bar{\Psi}^o_0\\
\phantom{\sigma'^o_2}=\bar{\sigma}^o\left(U^o_0+\frac{\Lambda}{6}|\sigma^o|^2\right)-\frac{\Lambda}{12}\bar{\Psi}^o_0\\
\rho'^o_2=\Psi^o_2+\sigma^o\dot{\bar{\sigma}}^o,\\
\textrm{in accordance to (10) as well.}
\end{cases}
\end{eqnarray}

(26) $D'\rho'-\delta\kappa'=\sigma'\bar{\sigma}'+\rho'(\rho'-2\textrm{Re}(\gamma))-2\alpha'\kappa'+k\phi_2\bar{\phi}_2$.
\begin{eqnarray}
r^2&:&0=0\\
r&:&0=0\\
1&:&0=0\\
r^{-1}&:&\dot{\rho}'^o_1=\frac{\Lambda}{3}\frac{\partial}{\partial u}(|\sigma^o|^2)+\frac{\Lambda}{3}\textrm{Re}(\eth\bar{\omega}^o_1),\textrm{ giving}\\
&\ &\rho'^o_1=\Theta(\theta, \phi)+\frac{\Lambda}{3}|\sigma^o|^2+\frac{\Lambda}{3}\int{\textrm{Re}(\eth\bar{\omega}^o_1)du}.
\end{eqnarray}
Here, $\Theta(\theta, \phi)$ is an arbitrary real function. From (25), this gives
\begin{eqnarray}
\dot{U}^o_0&=&-\frac{\Lambda}{2}\frac{\partial}{\partial u}(|\sigma^o|^2)-\frac{\Lambda}{3}\textrm{Re}(\eth\bar{\omega}^o_1),\\
U^o_0&=&-\Theta(\theta, \phi)-\frac{\Lambda}{2}|\sigma^o|^2-\frac{\Lambda}{3}\int{\textrm{Re}(\eth\bar{\omega}^o_1)du}.
\end{eqnarray}

(27) $D'\sigma'-\delta'\kappa'=\sigma'(2\textrm{Re}(\rho')-2\gamma-2i\textrm{Im}(\gamma))+2\alpha\kappa'+\Psi_4$.
\begin{eqnarray}
r&:&0=0\\
1&:&0=0\\
r^{-1}&:&\Psi^o_4=-\ddot{\bar{\sigma}}^o-\frac{2\Lambda}{3}\bar{\sigma}^oU^o_0-\frac{\Lambda}{3}\eth'\bar{\omega}^o_1-\frac{\Lambda^2}{9}\bar{\sigma}^o|\sigma^o|^2+\frac{\Lambda^2}{36}\bar{\Psi}^o_0\\
&\ &\phantom{\Psi^o_4}=-\ddot{\bar{\sigma}}^o+\frac{2\Lambda}{3}\bar{\sigma}^o\Theta-\frac{\Lambda}{3}\eth'\bar{\omega}^o_1+\frac{2\Lambda^2}{9}\bar{\sigma}^o|\sigma^o|^2+\frac{2\Lambda^2}{9}\bar{\sigma}^o\int{\textrm{Re}(\eth\bar{\omega}^o_1)du}+\frac{\Lambda^2}{36}\bar{\Psi}^o_0.\ \ \ \ \ 
\end{eqnarray}

(28) $D'\xi^\theta-\delta X^\theta=(\rho'+2i\textrm{Im}(\gamma))\xi^\theta+\bar{\sigma}'\overline{\xi^\theta}$.

(29) $D'\xi^\phi-\delta X^\phi=(\rho'+2i\textrm{Im}(\gamma))\xi^\phi+\bar{\sigma}'\overline{\xi^\phi}$.
\begin{eqnarray}
1&:&
\begin{cases}0=0\\
0=0
\end{cases}\\
r^{-1}&:&
\begin{cases}(\dot{\xi}^\theta)^o=-\frac{\Lambda}{3}\sigma^o(\overline{\xi^\theta})^o\label{xitheta}\\
(\dot{\xi}^\phi)^o=-\frac{\Lambda}{3}\sigma^o(\overline{\xi^\phi})^o\label{xiphi}
\end{cases}\\
r^{-2}&:&
\begin{cases}0=0\\
0=0
\end{cases}\\
r^{-3}&:&
\begin{cases}0=0\\
0=0.
\end{cases}
\end{eqnarray}

(30) $D'\alpha-\delta'\gamma=\alpha(\bar{\rho}'-2i\textrm{Im}(\gamma))+(\bar{\alpha}-2\alpha')\sigma'-\rho\kappa'-\Psi_3$.

(31) $D'\alpha'+\delta\gamma=\alpha'(\rho'+2i\textrm{Im}(\gamma))-\alpha\bar{\sigma}'+\sigma\kappa'-\tau\rho'+k\phi_1\bar{\phi}_2$.
\begin{eqnarray}
1&:&
\begin{cases}
0=0\\
0=0
\end{cases}\\
r^{-1}&:&
\begin{cases}
\dot{\alpha}^o=\frac{\Lambda}{3}\bar{\alpha}^o\bar{\sigma}^o+\frac{\Lambda}{6}\bar{\omega}^o_1\label{alpha}\\
0=0
\end{cases}\\
r^{-2}&:&
\begin{cases}0=0\\
0=0.
\end{cases}
\end{eqnarray}

(32) $\delta\rho-\delta'\sigma=(\bar{\alpha}-\alpha')\rho-(3\alpha+\bar{\alpha}')\sigma-\Psi_1+k\phi_0\bar{\phi}_1$.
\begin{eqnarray}
r^{-3}&:&\omega^o_1=\eth'\sigma^o\\
r^{-4}&:&0=0.
\end{eqnarray}

(33) $\delta'\rho'-\delta\sigma'=(\bar{\alpha}'-\alpha)\rho'-(3\alpha'+\bar{\alpha})\sigma'-\Psi_3+k\phi_2\bar{\phi}_1$.
\begin{eqnarray}
r^{-1}&:&0=0\\
r^{-2}&:&\Psi^o_3=-\eth\dot{\bar{\sigma}}^o-\frac{\Lambda}{6}\bar{\Psi}^o_1+\eth'U^o_0+\frac{\Lambda}{3}\sigma^o\eth'\bar{\sigma}^o.
\end{eqnarray}
Well,
\begin{eqnarray}
[\partial_u,\eth]\bar{\sigma}^o&=&\partial_u((\xi^\theta)^o\partial_\theta\bar{\sigma}^o+(\xi^\phi)^o\partial_\phi\bar{\sigma}^o-4\bar{\alpha}^o\bar{\sigma}^o)-\eth\dot{\bar{\sigma}}^o\label{com1}\\
&=&(\dot{\xi}^\theta)^o\partial_\theta\bar{\sigma}^o+(\dot{\xi}^\phi)^o\partial_\phi\bar{\sigma}^o-4\dot{\bar{\alpha}}^o\bar{\sigma}^o+\eth\dot{\bar{\sigma}}^o-\eth\dot{\bar{\sigma}}^o\\
&=&-\frac{\Lambda}{3}\sigma^o((\overline{\xi^\theta})^o\partial_\theta\bar{\sigma}^o+(\overline{\xi^\phi})^o\partial_\phi\bar{\sigma}^o+4\alpha^o\bar{\sigma}^o)-\frac{2\Lambda}{3}\bar{\sigma}^o\eth'\sigma^o\\
&=&-\frac{\Lambda}{3}\sigma^o\eth'\bar{\sigma}^o-\frac{2\Lambda}{3}\bar{\sigma}^o\eth'\sigma^o.\label{com2}
\end{eqnarray}
So,
\begin{eqnarray}
\Psi^o_3=-\partial_u(\eth\bar{\sigma}^o)-\frac{\Lambda}{6}\bar{\Psi}^o_1+\eth'U^o_0-\frac{2\Lambda}{3}\bar{\sigma}^o\eth'\sigma^o,
\end{eqnarray}
as expected from (17) and (22) in Eq. (\ref{Psithree}).

(34) $\textrm{Im}(\delta'\xi^\theta)=\textrm{Im}((\alpha+\bar{\alpha}')\xi^\theta)$.

(35) $\textrm{Im}(\delta'\xi^\phi)=\textrm{Im}((\alpha+\bar{\alpha}')\xi^\phi)$.
\begin{eqnarray}
r^{-2}&:&
\begin{cases}\textrm{Im}(\eth'(\xi^\theta)^o)=0\label{1}\\
\textrm{Im}(\eth'(\xi^\phi)^o)=0.
\end{cases}
\end{eqnarray}

(36) $\displaystyle \delta\alpha+\delta'\alpha'=\alpha\bar{\alpha}+\alpha'\bar{\alpha}'+2\alpha\alpha'-\rho\rho'+\sigma\sigma'-\Psi_2+\frac{\Lambda}{6}+k\phi_1\bar{\phi}_1$.
\begin{eqnarray}
1&:&0=0\\
r^{-1}&:&0=0\\
r^{-2}&:&2\textrm{Re}(\delta^o\alpha^o)-4|\alpha^o|^2=\Theta+\frac{\Lambda}{3}\int{\textrm{Re}(\eth^2\bar{\sigma}^o)du}.\label{Gauss}
\end{eqnarray}

(37) $(D'-2\gamma-\rho')\phi_0=(\delta-2\tau)\phi_1+\sigma\phi_2$.
\begin{eqnarray}
r^{-2}&:&0=0\\
r^{-3}&:&\dot{\phi}^o_0=\eth\phi^o_1+\sigma^o\phi^o_2+\frac{\Lambda}{6}\phi^1_0.
\end{eqnarray}

(38) $(D'-2\rho')\phi_1=(\delta-\tau-2\alpha')\phi_2-\kappa'\phi_0$.
\begin{eqnarray}
r^{-1}&:&0=0\\
r^{-2}&:&\dot{\phi}^o_1=\eth\phi^o_2-\frac{\Lambda}{6}\eth'\phi^o_0.
\end{eqnarray}

(39) $(D'-4\gamma-\rho')\Psi_0-(\delta-4\tau+2\alpha')\Psi_1-3\sigma\Psi_2=k\bar{\phi}_1(\delta+2\alpha')\phi_0-k\bar{\phi}_2D\phi_0+2k\sigma\phi_1\bar{\phi}_1$.
\begin{eqnarray}
r^{-4}&:&0=0\\
r^{-5}&:&\dot{\Psi}^o_0=\eth\Psi^o_1+3\sigma^o\Psi^o_2+3k\phi^o_0\bar{\phi}^o_2+\frac{\Lambda}{6}\Psi^1_0.
\end{eqnarray}

(40) $\kappa'\Psi_0+(D'-2\gamma-2\rho')\Psi_1-(\delta-3\tau)\Psi_2-2\sigma\Psi_3=k\bar{\phi}_1(D'-2\gamma)\phi_0-k\bar{\phi}_2(\delta'-2\alpha)\phi_0+2k\phi_1(\tau\bar{\phi}_1-\rho\bar{\phi}_2)$.
\begin{eqnarray}
r^{-3}&:&0=0\\
r^{-4}&:&\dot{\Psi}^o_1=\eth\Psi^o_2+2\sigma^o\Psi^o_3+2k\phi^o_1\bar{\phi}^o_2-\frac{\Lambda}{6}\eth'\Psi^o_0+\frac{k\Lambda}{3}\phi^o_0\bar{\phi}^o_1.
\end{eqnarray}

(41) $2\kappa'\Psi_1+(D'-3\rho')\Psi_2-(\delta-2\tau-2\alpha')\Psi_3-\sigma\Psi_4=k\bar{\phi}_1(\delta-2\alpha')\phi_2-k\bar{\phi}_2D\phi_2+2k\rho'\phi_1\bar{\phi}_1$.
\begin{eqnarray}
r^{-2}&:&0=0\\
r^{-3}&:&-\partial_u(\Psi^o_2+\sigma^o\dot{\bar{\sigma}}^o)=-|\dot{\sigma}^o|^2-\eth\Psi^o_3-k|\phi^o_2|^2-\frac{2\Lambda}{3}\Theta|\sigma^o|^2+\frac{\Lambda}{3}\sigma^o\eth'\eth\bar{\sigma}^o+\frac{\Lambda}{6}\eth'\Psi^o_1\nonumber\\&\ &\phantom{-\partial_u(\dot{\Psi}^o_2+\sigma^o\bar{\sigma}^o)=}-\frac{2\Lambda^2}{9}|\sigma^o|^4-\frac{2\Lambda^2}{9}|\sigma^o|^2\int{\textrm{Re}(\eth^2\bar{\sigma}^o)du}-\frac{\Lambda^2}{18}\textrm{Re}(\bar{\sigma}^o\Psi^o_0)+\frac{k\Lambda^2}{36}|\phi^o_0|^2.\nonumber\\
\end{eqnarray}

(42) $3\kappa'\Psi_2+(D'+2\gamma-4\rho')\Psi_3-(\delta-\tau-4\alpha')\Psi_4=k\bar{\phi}_1(D'+2\gamma)\phi_2-k\bar{\phi}_2(\delta'+2\alpha)\phi_2+2k\phi_1(\kappa'\bar{\phi}_1-\sigma'\bar{\phi}_2)$.
\begin{eqnarray}
r^{-1}&:&0=0\\
r^{-2}&:&0=0.\label{above}
\end{eqnarray}
In showing that Eq. (\ref{above}) is identically satisfied, we need the following four commutators:
\begin{eqnarray}
[\partial_u,\eth']U^o_0&=&-\frac{\Lambda}{3}\bar{\sigma}^o\eth U^o_0\\
{[}\partial_u,\eth]\dot{\bar{\sigma}}^o&=&-\frac{\Lambda}{3}\sigma^o\eth'\dot{\bar{\sigma}}^o-\frac{2\Lambda}{3}\dot{\bar{\sigma}}^o\eth'\sigma^o\\
{[}\partial_u,\eth']\bar{\sigma}^o&=&\frac{\Lambda}{3}\bar{\sigma}^o\eth\bar{\sigma}^o\\
{[}\eth,\eth']\eth\bar{\sigma}^o&=&2(2\textrm{Re}(\delta^o\alpha^o)-4|\alpha^o|^2)\eth\bar{\sigma}^o\\
&=&2\left(\Theta+\frac{\Lambda}{3}\int{\textrm{Re}(\eth^2\bar{\sigma}^o)du}\right)\eth\bar{\sigma}^o.
\end{eqnarray}
The derivation of the first three of these commutators is similar to that for $[\partial_u,\eth]\bar{\sigma}^o$, as was done in (33) (Eqs. (\ref{com1})-(\ref{com2})). The derivation of the $[\eth,\eth']\eth\bar{\sigma}^o$ commutator involves (34) and (35), viz. Eq. (\ref{1}) \footnote{This is consistent with (4.14.1) in Penrose-Rindler \cite{Pen87} for $[\eth,\eth']\eth\bar{\sigma}^o$, where $\eth\bar{\sigma}^o$ is a quantity with $p=-2$ and $q=0$. The term involving $\textrm{Im}(\rho')$\th\ is zero because $\gamma'=0$ gives \th$\ =\partial/\partial r$, but $\eth\bar{\sigma}^o$ is independent of $r$. The term involving $\textrm{Im}(\rho)$\th$'$ is zero because $\rho$ is real.}. Note that all these commutators (including the one used in (33)) are zero when $\Lambda=0$, except $[\eth,\eth']\eth\bar{\sigma}^o=2\Theta\eth\bar{\sigma}^o$.

In the case of de Sitter spacetime where $\sigma^o=0$, we have $U^o_0=-1/2$. Hence we take $\Theta=1/2$.

\bibliographystyle{spphys}       
\bibliography{Citation}

\end{document}